\documentclass[aps,prb,reprint,superscriptaddress,twocolumn, longbibliography]{revtex4-2}

\pdfoutput=1

\usepackage{graphicx}
\usepackage{amsmath,amsfonts,amssymb,mathtools}
\usepackage{color}
\usepackage{subfigure}
\usepackage{physics}
\usepackage{dsfont}
\usepackage{hyperref}
\usepackage{booktabs}
\usepackage{url}
\usepackage{leftidx}
\usepackage{textcomp}
\usepackage{gensymb}
\usepackage{threeparttable}
\usepackage{rotating}
\usepackage{esvect}
\usepackage{booktabs}
\usepackage{marvosym}
\usepackage{lmodern}
\usepackage[ruled,vlined]{algorithm2e}
\usepackage[noend]{algpseudocode}
\usepackage[dvipsnames]{xcolor}

\definecolor{emerald}{rgb}{0.31, 0.78, 0.47}
\definecolor{blue(ncs)}{rgb}{0.0, 0.53, 0.74}

\hypersetup{
     colorlinks=true,
     linkcolor=magenta,
     filecolor=blue,
     citecolor=emerald,      
     urlcolor =blue(ncs),
}

\usepackage{chngcntr}

\DeclareMathAlphabet{\pazocal}{OMS}{zplm}{m}{n}

\newcommand{\bo}[1]{\boldsymbol{#1}}

\newcommand{\D}{\mathrm{d}}
\newcommand{\f}[2]{\frac{#1}{#2}}

\begin{document}

\title{Heat capacity double transitions in time-reversal symmetry broken superconductors}

\author{Henrik S.~R{\o}ising}
\email{henrik.roising@nbi.ku.dk}
\affiliation{Niels Bohr Institute, University of Copenhagen, DK-2200 Copenhagen, Denmark}

\author{Glenn Wagner}
\affiliation{Department of Physics, University of Zurich, Winterthurerstrasse 190, 8057 Zurich, Switzerland}

\author{Merc{\`e} Roig}
\affiliation{Niels Bohr Institute, University of Copenhagen, DK-2200 Copenhagen, Denmark}

\author{Astrid T.~R{\o}mer}
\affiliation{Niels Bohr Institute, University of Copenhagen, DK-2200 Copenhagen, Denmark}
  \affiliation{Danish Fundamental Metrology, Kogle All\'e 5, 2970 H{\o}rsholm, Denmark}
  
\author{Brian M.~Andersen}
\affiliation{Niels Bohr Institute, University of Copenhagen, DK-2200 Copenhagen, Denmark}  
 
\date{\today}

\begin{abstract}
Standard superconductors display a ubiquitous discontinuous jump in the electronic specific heat at the critical superconducting transition temperature. In a growing class of unconventional superconductors, however, a second order parameter component may get stabilized and produce a second heat capacity jump at a lower temperature, typically associated with the spontaneous breaking of time-reversal symmetry. The splitting of the two specific heat discontinuities can be controlled by external perturbations such as chemical substitution, hydrostatic pressure, or uniaxial strain. We develop a theoretical quantitative multi-band framework to determine the ratio of the heat capacity jumps, given the band structure and the order parameter momentum structure. We discuss the conditions of the gap profile which determine the amplitude of the second jump. We apply our formalism to the case of Sr$_2$RuO$_4$, and using the gap functions from a microscopic random phase approximation calculation, we show that  recently-proposed accidentally degenerate order parameters may exhibit a strongly suppressed second heat capacity jump. We discuss the origin of this result and consider also the role of spatial inhomogeneity on the specific heat. Our results provide a possible explanation of why a second heat capacity jump has so far evaded experimental detection in Sr$_2$RuO$_4$.  
\end{abstract}

\maketitle

%
\section{Introduction}
%
A growing number of superconductors realize multi-component condensates, some with clear evidence for spontaneous time-reversal symmetry breaking (TRSB) below the critical superconducting transition temperature $T_c$~\cite{Ghosh2EA20}. Such states of matter exhibit persistent super-currents at material edges, dislocations, or around nonmagnetic defect sites~\cite{Lee2009,Clara2022}. The term ``multi-component'' refers to real or complex superpositions of two allowed symmetry-distinct pairing symmetries. An important bulk thermodynamic probe which can determine the associated phase transitions is specific heat measurements. Possible multi-component superconductors of current interest include several uranium-based materials UPt$_3$~\cite{StewartEA84, JoyntEA02}, UTe$_2$~\cite{RanEA19, AokiEA22}, URu$_2$Si$_2$~\cite{SchlabitzEA86, MydoshEA11}, but also compounds like Sr$_2$RuO$_4$~\cite{MaenoEA94, MackenzieEA17}, Ba$_{1-x}$K$_x$Fe$_2$As$_2$~\cite{RotterEA08, BokerEA17}, LaNiC$_2$~\cite{HillierEA09}, and PrOs$_4$Sb$_{12}$~\cite{BauerEA02, AokiEA03}. Here, particularly UPt$_3$ is known to exhibit a clear double transition possibly into an exotic low-temperature spin-triplet $f + i f$ superconducting state~\cite{JoyntEA02}. At present, the nature of superconductivity in UTe$_2$ and Sr$_2$RuO$_4$ is controversial and of high interest to the community. In this respect, the recent discovery of a large Knight shift suppression~\cite{PustogowEA19, ChronisterEA20} in the superconducting state of Sr$_2$RuO$_4$ has ignited tremendous renewed interest in the exploration of the superconducting state of this material. Cooper pairing in Sr$_2$RuO$_4$ is not of predominant spin-triplet character, but rather of the spin-singlet form. The stakes are significant, not just because of a desire to explain superconductivity in this particular compound, but also because it challenges our ability to experimentally determine, and theoretically describe, unconventional superconductivity arising from many-body instabilities in Fermi liquids. At present, it is unclear whether the standard framework for unconventional superconductivity in terms of repulsive short-range Coulomb interactions properly captures the pairing structure of Sr$_2$RuO$_4$. The theoretical description is complicated by a large spin-orbit coupling and a multi-band (multi-orbital) Fermi surface, which likely supports several near-degenerate symmetry-distinct superconducting phases~\cite{ScaffidiEA14, RoisingEA19, RomerEA19, Scaffidi20, WangEA22}.

The current status of superconductivity in Sr$_2$RuO$_4$ can be condensed into the following open question: what nodal spin-singlet time-reversal symmetry breaking superconducting multi-component order parameter, whose product transforms as $B_{2g}$, is realized below $T_c$? As mentioned above, nuclear magnetic resonance (NMR) dictates a condensate of primarily spin-singlet nature, implying even-parity Cooper pairs. Spectroscopic probes~\cite{FirmoEA13, MadhavenEA19}, specific heat~\cite{NishiZakiEA00, LiEA19}, and thermal conductivity measurements~\cite{HassingerEA17} are consistent with nodal quasiparticle excitations. Optical Kerr rotation measurements~\cite{XiaEA06} and muon spin relaxation ($\mu$SR)~\cite{Luke98} find evidence for TRSB below $T_c$, and ultrasound experiments also point to multi-component condensates~\cite{GhoshEA20, Benhabib2021}. Specifically, the presence of a discontinuity in the shear elastic modulus $c_{66}$ (the $B_{2g}$ channel) at $T_c$ seems only consistent with situations where the product of the two involved order parameters transforms as the $B_{2g}$ irreducible representation (irrep) of the point group $D_{4h}$ relevant for Sr$_2$RuO$_4$~\cite{GhoshEA20, Benhabib2021}. Thus at face value these constraints allow only  $d_{xz}+id_{yz}$~\cite{ZuticEA05, SuhEA19}, $s’ + i d_{xy}$~\cite{RomerEA21, WangEA22}, and $d_{x^2-y^2}+ig_{xy(x^2-y^2)}$~\cite{KivelsonEA20, WillaEA21} orders. [We denote by $s’$ an order parameter with a nodal spectrum and $A_{1g}$ symmetry. We note that nodal quasiparticles for the $s'+id_{xy}$ scenario require matching of accidental nodes for the two components.] Of these candidates the former is attractive due to its inherent 2D irrep character, but appears unlikely from its nodal properties and 3D nature~\cite{RoisingEA19, RomerEA22}. In addition, $d_{xz}+id_{yz}$ implies a coupling to the $(c_{11}-c_{12})/2$ ($B_{1g}$) shear channel, which has not been observed. On the other hand, the latter two candidates currently appear consistent with all the recent developments summarized above, but require fortuitous degeneracy between $s’$ and $d_{xy}$, or $d_{x^2-y^2}$ and $g_{xy(x^2-y^2)}$ pairing symmetries.

For all these mentioned multi-component pairing candidates, however, there is an important “fly in the ointment” that needs to be addressed. Specific heat measurements under uniaxial pressure up to around $0.6$ GPa detect only a single phase transition even as the leading $T_c$ rises from 1.5 to 3.5 K~\cite{LiEA19}. This points to a superconducting phase consisting of a single order parameter. Nevertheless, pursuing the multi-component candidates further, the naive expectation of such a symmetry-breaking perturbation would be a split transition with two associated distinguishable peaks in the specific heat versus temperature. Indeed this particular issue has given rise to alternative scenarios for Sr$_2$RuO$_4$ based on inhomogeneous phases with secondary order parameter components stabilized only near edge dislocations of the crystal structure~\cite{WillaEA21}. Alternatively, the second transition may simply be very weak and so far indiscernible experimentally.  According to Ref.~\onlinecite{LiEA19}, this requires that the second transition be less than $\sim 5\%$ of the primary phase transition discontinuity. Previous theoretical investigations of the size of the second specific heat jump were reported in Ref.~\onlinecite{WagnerEA21} where a microscopically-derived Ginzburg--Landau approach was taken to numerically explore the specific heat of mainly a $d_{x^2-y^2}+ig_{xy(x^2-y^2)}$ solution in the unstrained case.

Here, we perform a general theoretical investigation of the specific heat transition of double-component superconducting phase transitions, and identify the key conditions determining the ratio of the amplitudes of the specific heat jumps at the two transition temperatures. This study is of relevance to all TRSB two-component superconducting phase transitions. In addition, we compute the specific heat explicitly from recent microscopically obtained  $s’ + i d_{xy}$, and $d_{x^2-y^2}+ig_{xy(x^2-y^2)}$ superconducting orders specific to Sr$_2$RuO$_4$, and consider the effects of uniaxial strain and spatial inhomogeneity. We discuss the results in light of the recent specific heat measurements on Sr$_2$RuO$_4$, and conclude that the fortuitously degenerate superconducting candidate orders obtained from realistic material-specific models are consistent with currently available specific heat data.

The paper is organized as follows. In Sec.~\ref{sec:HeatCapacityMain} we derive the ratio of the heat capacity jumps for a general two-component order parameter. The result is explored within a microscopic Ginzburg--Landau framework, and with a numerical demonstration in a simple one-band model. In Sec.~\ref{sec:SRO} we apply our formalism to the case of Sr$_2$RuO$_4$ by calculating the heat capacity for two recently-proposed order parameter candidates as a function of uniaxial in-plane strain. In Sec.~\ref{sec:OtherMaterials} we discuss a series of other putative time-reversal symmetry-broken superconductors with possible double transitions in light of our results at the qualitative level. The paper is concluded in Sec.~\ref{sec:Conclusions}. In Appendix~\ref{sec:HeatCapacity} we provide details of the derivation of the heat capacity in multiband superconductors. Appendix~\ref{sec:GL} includes a self-contained derivation of the microscopic Ginzburg--Landau approach. In Appendix~\ref{sec:Bandstructure} we list the three-band tight-binding model of Sr$_2$RuO$_4$ and its strain parametrization, and in Appendix~\ref{sec:MorePlots} we provide supplementary heat capacity plots for Sr$_2$RuO$_4$.

%
\section{Heat capacity of multicomponent superconductors}
\label{sec:HeatCapacityMain}
%
We consider a two-component order parameter with a general temperature dependence, as parameterized by the two functions $\Delta_0(T)$ and $X(T)$:
\begin{equation}
\Delta(T, \bo{p}) = \Delta_0(T) \left[ f_{a}(\bo{p}) + i X(T) f_{b}(\bo{p}) \right].
\label{eq:OPansatz}
\end{equation}
Here, $a$ and $b$ are symmetry labels denoting two different one-dimensional irreducible representations (irreps) of the appropriate point group. The order parameter is in general multiband; a band index has been left implicit in the above ansatz. The form factors are normalized as $\max_{\bo{p}\in \mathrm{FS}} \lvert f_{a/b}(\bo{p}) \rvert = 1$. 

\begin{figure}[tb]
	\centering
	\includegraphics[width=0.85\linewidth]{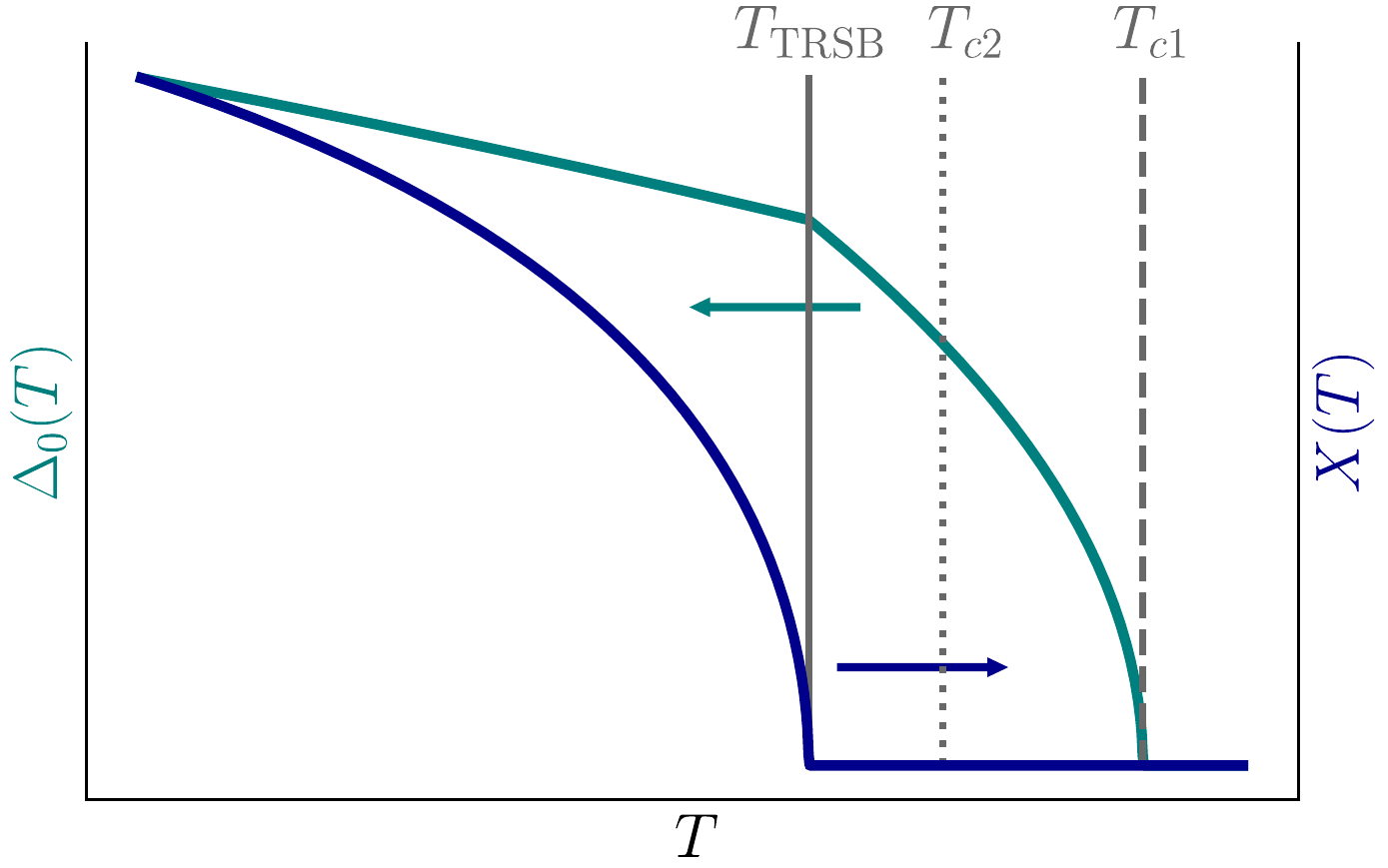}
\caption{Sketch of the order parameter temperature dependence, where $\Delta(T, \bo{p}) = \Delta_0(T) \left[ f_{a}(\bo{p}) + i X(T) f_{b}(\bo{p}) \right]$. When the two components have critical temperatures $T_c \coloneqq T_{c1}$ and $T_{c2}$, repulsion leads to the onset of the second component at $T_{\mathrm{TRSB}} < T_{c2}$, for which time-reversal symmetry is broken (blue line). Due to repulsion $\Delta_0(T)$ (green line) becomes non-analytic at $T_{\mathrm{TRSB}}$. The non-analyticity leads to a reduction of the second heat capacity jump.}
	\label{fig:Condensation}
\end{figure}

In the current scenario where the two components of the condensate belong to different irreps, the temperatures at which $\Delta_0(T)$ and $X(T)$ onset, which we denote by $T_c$ and $T_{\mathrm{TRSB}}$, respectively, are generally different. Generically, re-condensation at $T_{\mathrm{TRSB}}$ makes $\Delta_0(T)$ non-analytic at this temperature, which has important consequences for the second specific heat jump that we address in detail in the following. Sketches of the temperature dependencies of $\Delta_0$ and $X$ are shown in Fig.~\ref{fig:Condensation}.

\subsection{The heat capacity discontinuities}
The heat capacity of a superconductor, $C(T) = T \partial S / \partial T$, is derived from the entropy of the free Fermi-gas~\cite{Tinkham75} (see also Appendix~\ref{sec:HeatCapacity} for a detailed derivation)
\begin{equation}
\begin{aligned}
S &= -2 \sum_{\mu, \bo{k}} \Big[ n_F[E_{\mu}(\bo{k})] \ln n_F[E_{\mu}(\bo{k})] \\
&\hspace{30pt} + \big(1-n_F[E_{\mu}(\bo{k})] \big) \ln\big( 1 - n_F[E_{\mu}(\bo{k})] \big) \Big],
\end{aligned}
\label{eq:EntropyMain}
\end{equation}
where $n_F(E)$ denotes the Fermi function and $E_{\mu}(\bo{k})$ is the quasiparticle excitation energy of band $\mu$ at momentum $\bo{k}$. As the electrons condense into Cooper pairs, both at the primary ($T_c$) and the secondary ($T_{\mathrm{TRSB}}$) transitions, there is an associated reduction of entropy that results in a discontinuous jump in the heat capacity. We define the heat capacity jumps as the change in value across the respective transitions, $\Delta C_{\mathrm{SC}}/T_c \coloneqq [\lim_{T \to T_c^-} C(T) - \lim_{T \to T_c^+} C(T)]/T_c$ and $\Delta C_{\mathrm{TRSB}}/T_{\mathrm{TRSB}} \coloneqq [\lim_{T \to T_{\mathrm{TRSB}}^-} C(T) - \lim_{T \to T_{\mathrm{TRSB}}^+} C(T)]/T_{\mathrm{TRSB}}$. It is a straightforward task to calculate the ratio of the two jumps in the thermodynamic limit, which we denote by $\eta$, starting from Eq.~\eqref{eq:EntropyMain} and using the general order parameter ansatz of Eq.~\eqref{eq:OPansatz}. We provide the details of this calculation in  Appendix~\ref{sec:HeatCapacity} and summarize the main result here. When $\eta$ is expressed as a Fermi surface average, the final result reads
\begin{widetext}
\begin{equation}
\eta = \frac{\Delta C_{\mathrm{TRSB}} / T_{\mathrm{TRSB}} }{\Delta C_{\mathrm{SC}} / T_c } = \frac{T_c}{T_{\mathrm{TRSB}}} \Big[  A \frac{\langle \lvert f_a(\bo{p}) \rvert^2 \pazocal{I}[z(\bo{p})] \rangle_{\mathrm{FS}}}{ \langle \lvert f_a(\bo{p}) \rvert^2 \rangle_{\mathrm{FS}}} + B \frac{\langle \lvert f_b(\bo{p}) \rvert^2 \pazocal{I}[z(\bo{p})] \rangle_{\mathrm{FS}}}{ \langle \lvert f_a(\bo{p}) \rvert^2 \rangle_{\mathrm{FS}}} \Big].
\label{eq:EtaMain}
\end{equation}
\end{widetext}
This result comes about as two terms are picked up in the evaluation of $\Delta C_{\mathrm{TRSB}}/T_{\mathrm{TRSB}}$ since generally the temperature dependence of both order parameter components, $\Delta_0^2(T)$ and $\Delta_0^2(T)X^2(T)$, are non-analytic at $T_{\mathrm{TRSB}}$, as caused by terms in the free energy that couple the two components. Formally, the non-analyticity is expressed via
\begin{align}
A &= Q(\Delta_0,T_{\mathrm{TRSB}})/Q(\Delta_0, T_c), \label{eq:Ageneral} \\
B &= Q(\Delta_0 X,T_{\mathrm{TRSB}})/Q(\Delta_0, T_c), \label{eq:Bgeneral} \\
Q(g,T_0) &\coloneqq  \lim_{T \to T_0^-} \frac{\partial g^2(T)}{\partial T} - \lim_{T \to T_0^+} \frac{\partial g^2(T)}{\partial T}. \label{eq:Qfunc}
\end{align}
Finally, in Eq.~\eqref{eq:EtaMain} we introduced the dimensionless integral $\pazocal{I}(z) \coloneqq \int_{0}^{\infty} \D u~\cosh^{-2}\left(\sqrt{u^2+z^2}\right)$ with $z(\bo{p}) = \frac{\Delta_0(T_{\mathrm{TRSB}})}{2T_{\mathrm{TRSB}}} f_a(\bo{p})$, and the Fermi surface averages are evaluated as $\langle f \rangle_{\mathrm{FS}} \coloneqq \sum_{\mu} \int_{S_{\mu}} \frac{\D \bo{k}}{(2\pi)^d} \frac{f}{v_{\mu}(\bo{k})}$, with $v_{\mu}(\bo{k}) = \lvert \nabla \xi_{\mu}(\bo{k}) \rvert$ being the Fermi velocity of band $\mu$.

In the particular case of $\Delta_0(T)$ being analytic at $T_{\mathrm{TRSB}}$, we see that $A = 0$. In the BCS-like case of $\Delta_0(T) = \Delta_{0a} (1-T/T_c)^{1/2}$, and $X(T) = \Delta_{0b}/\Delta_{0a} [(1-T/T_{\mathrm{TRSB}})/(1-T/T_c)]^{1/2}$ we recover the result of Ref.~\onlinecite{KivelsonEA20}, i.e.,
\begin{equation}
    \eta_0 = \left(\frac{T_c}{T_{\mathrm{TRSB}}}\right)^2 \left(\frac{\Delta_{0b}}{\Delta_{0a}}\right)^2 \frac{\langle \lvert f_b(\bo{p}) \rvert^2 \pazocal{I}[z(\bo{p})] \rangle_{\mathrm{FS}}}{\langle \lvert f_a(\bo{p}) \rvert^2 \rangle_{\mathrm{FS}} }.
    \label{eq:eta0}
\end{equation}
We stress that when the two order parameter components repel, the resulting non-analyticity of $\Delta_0(T)$ suppresses the second heat capacity jump, with the conditions for an observable second transition being quantitatively dictated by Eq.~\eqref{eq:EtaMain}.

\subsection{Ginzburg--Landau theory}
\label{sec:GLtheory}
To shed light on the heat capacity jump ratio in Eq.~\eqref{eq:EtaMain}, which involves the temperature dependence of the two order parameter components, and to facilitate a calculation of the coefficients $A$ and $B$, we turn to a microscopic Ginzburg--Landau theory building on seminal work by Gor'kov~\cite{Gorkov59}. For a multi-component order parameter we derive the general quartic order free energy in the superconducting phase in Appendix~\ref{sec:GL}, with further details given in Ref.~\onlinecite{WagnerEA21}. In the two-component case of Eq.~\eqref{eq:OPansatz}, when explicitly including band indices $\mu$, the free energy takes the form 
\begin{equation}
\begin{aligned}
    \Delta F &= \Delta_0^2(T) \bigg[\alpha_{a}(T,T_{c1}) + \alpha_{b}(T,T_{c2})X^2(T)\bigg] \\ 
    &\hspace{20pt} + \Delta_0^4(T)\bigg[\beta_{a} +2\beta_{ab} X^2(T) + \beta_{b} X^4(T)\bigg], \label{eq:FreeEnergy}
\end{aligned}
\end{equation}
where
\begin{align}
\alpha_{j}&(T,T_{cj}) = - V \sum_{\mu}\int \f{\D \bo{p}}{(2\pi)^d}~\Big( \f{\tanh\left[ \xi_{\mu}(\bo{p}) / (2T) \right] }{2\xi_{\mu}(\bo{p})} \nonumber \\
&\hspace{50pt} -\f{\tanh\left[ \xi_{\mu}(\bo{p}) / (2T_{cj}) \right]}{2\xi_{\mu}(\bo{p})} \Big) f_{j\mu}^2(\bo{p}) \label{eq:Alpha}, \\
\beta_{j} &= \sum_\mu\f{V}{2T_{c1}^3} \int \f{\D \bo{p}}{(2\pi)^d}~h(\xi_{\mu}(\bo{p})/ T_{c1} )f_{j\mu}^4(\bo{p}) \label{eq:Betaj}, \\
\beta_{ab} &= \sum_\mu\f{V}{2T_{c1}^3} \int \f{\D \bo{p}}{(2\pi)^d}~h( \xi_{\mu}(\bo{p})/ T_{c1} )f_{a\mu}^2(\bo{p})f_{b\mu}^2(\bo{p}) \label{eq:Betaab},
\end{align}
with $V$ being the reciprocal space volume, $j = a,~b$, and $h(x) \geq 0$ is sharply peaked around $x=0$ for low temperatures and is given in Eq.~\eqref{eq:hfunc}. Above, we asserted that $f_a$ condenses at $T_{c1}$ and $f_b$ at $T_{c2}$. 

The free energy is minimized over $\Delta_0(T)$ and $X(T)$, and in Appendix~\ref{sec:TwoComponent} we show that this results directly in $T_c = T_{c1}$ and $T_{\mathrm{TRSB}} = T_{\mathrm{TRSB}}(T_{c1}, T_{c2}) \leq T_{c2}$ (equality when $\beta_{ab} = 0$). Since the temperature dependence of the quartic order is weak and not a leading-order effect, we evaluate the $\beta$'s at $T = T_{c}$. When further using $\alpha_j(T,T_{cj}) = \alpha_j^0(T/T_{cj}-1)$ with $\alpha_j^0 \geq 0$, which holds close to the respective transition temperatures, we find that the Ginzburg--Landau solution yields 
\begin{align}
    A &= -\frac{\beta_{ab}}{\beta_{a}\beta_{b}-\beta_{ab}^2} \left[\beta_{ab} + \frac{\alpha_b^0}{\alpha_a^0}\frac{T_c}{T_{\mathrm{TRSB}}} \beta_{a} \right], \label{eq:CoeffA} \\
    B &= \frac{\beta_{a}}{\beta_{a}\beta_{b}-\beta_{ab}^2} \left[ \frac{\alpha_b^0}{\alpha_a^0}\frac{T_c}{T_{\mathrm{TRSB}}} \beta_{a} - \beta_{ab} \right], \label{eq:CoeffB}
\end{align}
for the coefficients in Eq.~\eqref{eq:EtaMain}. Owing to H{\"o}lder's inequality it follows that $\beta_{a}\beta_{b}-\beta_{ab}^2 \geq 0$. Moreover, since $\beta_{b},~\beta_{a} \geq 0$, we conclude that $A \leq 0$ and that $B$ in principle can take both signs but is only physically relevant in the positive semi-definite regime to yield an overall positive $\eta$.

It is worth noting that the free energy of Eq.~\eqref{eq:FreeEnergy}, derived in the case where the two order parameter components belong to different one-dimensional irreps of $D_{4}$, still holds under uniaxial in-plane strain where the crystal symmetries are reduced, $D_{4} \rightarrow D_{2}$ (note that $D_{2}$ only permits one-dimensional irreps), cf.~Ref.~\onlinecite{ThomasEA20}. Strain modifies both the band structure $\xi_{\mu}$ and the form factors $f_{a\mu}(\bo{p})$, but as long as the form factors are labelled appropriately by the irreps of the symmetry-reduced point group $D_{2}$, the coefficients in Eqs.~\eqref{eq:Alpha}-\eqref{eq:Betaab} still apply. 

\subsection{Parameters controlling the magnitude of the second jump}
\label{sec:Parameters}
The expression of Eq.~\eqref{eq:EtaMain}, along with the Ginzburg--Landau result of Eqs.~\eqref{eq:CoeffA} and \eqref{eq:CoeffB}, reveal the key knobs that dictate the size of the second heat capacity jump. 

First, $\eta$ is generally highly sensitive to the order of which the order parameter components onset. Intuitively, the more of the Fermi surface that is gapped-out by the primary component (at $T_{c}$), or correspondingly the fewer the quasiparticles left to be gapped-out by the secondary component (at $T_{\mathrm{TRSB}}$), the smaller the second jump. This can be inferred directly from Eq.~\eqref{eq:EtaMain} by the fact that $\langle \lvert f_a(\bo{p}) \rvert^2 \rangle_{\mathrm{FS}}$ appears in the denominators, so the order of the components is decisive for the size of $\eta$.

Second, $\eta$ is suppressed in cases where the two components share momentum structure over the Fermi surface. This can be seen from the behaviour of the dimensionless integral appearing in Eq.~\eqref{eq:EtaMain}, $\pazocal{I}(z) \approx \exp(- 7 \zeta(3) z^2 / \pi^2)$ (accurate to $\pazocal{O}(z^2)$). The second and dominant term in Eq.~\eqref{eq:EtaMain} is therefore approximately proportional to $\langle \lvert f_b(\bo{p}) \rvert^2 \exp(-c \lvert f_a(\bo{p}) \rvert^2) \rangle_{\mathrm{FS}}$, where $c$ is a positive constant. This expression will be small if, for a given momentum on the Fermi surface, $\lvert f_a(\bo{k}) \rvert$ is maximal (minimal) where $\lvert f_b(\bo{k}) \rvert$ is maximal (minimal). Consequentially, if the two components have a similar momentum structure over the Fermi surface, the second heat capacity jump is suppressed. In cases where the Fermi velocity is anisotropic, the regions with minimal velocity (particularly in the vicinity of van Hove points) will dominate the Fermi surface average and dictate where a matching momentum structure is decisive to get a suppressed second jump. We note that this criterion of a matching momentum structure is generally not limited to nodal regions, which is commonly accepted, but applies in principle to the entire Fermi surface.

Third, but indirectly related to the second point, increasing the repulsion between the components reduces the second jump. The repulsion is controlled by $\beta_{ab}$, which $A$ (in Eq.~\eqref{eq:CoeffA}) is directly proportional to. This is related to the second point because $\beta_{ab}$ is maximized when the components $f_a(\bo{p})$ and $f_b(\bo{p})$ share momentum structure. Technically, the closer one is to saturating H{\"o}lder's inequality $\beta_{ab}^2 \leq \beta_{a}\beta_{b}$, the smaller the second jump. 

Since $a$ and $b$ label different one-dimensional irreps both symmetry-breaking and symmetry-preserving perturbations (with respect to the point group), commonly labelled $\epsilon$, will \emph{a priori} cause a splitting of $T_c$ and $T_{\mathrm{TRSB}}$ since $\partial T_{c1} / \partial \epsilon \neq \partial T_{c2} / \partial \epsilon$ generally, and the ratio $T_c / T_{\mathrm{TRSB}}$ largely controls $\eta$. In the limit $\frac{T_c}{T_{\mathrm{TRSB}}} \gg \frac{\alpha_a^0}{\alpha_b^0} \frac{\beta_{ab}}{\beta_a}$ Eqs.~\eqref{eq:CoeffA} and \eqref{eq:CoeffB} tell us that $\eta \propto (T_c / T_{\mathrm{TRSB}})^2$.

Experimentally, natural symmetry-preserving perturbations to consider include, e.g.,~disorder, impurity doping~\cite{BastianSigrist20}, and hydrostatic pressure. In the case of Sr$_2$RuO$_4$, La substitution of Sr, taking the effective role of added disorder~\cite{ShenEA07}, and moderate hydrostatic pressure were applied, both resulting in no detectable temperature splitting~\cite{GrinenkoEA21}. In this case, it would also be of interest to see the result of Co and Mn substitution of Ru since this has been shown to enhance ferromagnetic and antiferromagnetic fluctuations, respectively~\cite{OrtmannEA13, ShengEA21}. Symmetry-breaking perturbations of experimental relevance include strain, which fundamentally decomposes into five types in the case of $D_{4}$: two compressional modes preserve the crystal symmetries, and three shear modes reduce the crystal symmetries~\cite{GhoshEA20}. In Sr$_2$RuO$_4$ the (strikingly different) effects of both symmetry-breaking (100) strain~\cite{SteppkeEA17} and symmetry-preserving (001) strain~\cite{JerzembeckEA21} have been studied.

\subsection{Demonstration: circular Fermi surface}
\label{sec:SingleBand}
In this section we demonstrate the basic principles stated in Sec.~\ref{sec:Parameters} for controlling the size of the second specific heat jump by calculating the heat capacity discontinuity from Eq.~\eqref{eq:EtaMain} in the simplest case of a circular Fermi surface with $\Delta_0(T) = \Delta_{0a} (1-T/T_c)^{1/2}$, and $X(T) = \Delta_{0b}/\Delta_{0a} [(1-T/T_{\mathrm{TRSB}})/(1-T/T_c)]^{1/2}$. The circular Fermi surface emerges, for example, on the square lattice with nearest-neighbor hopping, $\xi(\bo{k}) = -2t(\cos{k_x}+\cos{k_y}) - \mu$, in the low-filling limit $\mu \approx -4t$: $k_x^2+k_y^2 = k_F^2$, with $k_F = \sqrt{4+\mu/t}$~\cite{RomerEA15}. In this case the Fermi velocity, Fermi surface area, and density of states simplify to $v_F = 2 t k_F$, $\lvert S \rvert = 2\pi k_F$, and $\rho_0 = k_F / (4\pi t)$, respectively. The Fermi surface averages reduce to $\langle f \rangle_{\mathrm{FS}} = \frac{1}{2\pi k_F} \int_{0}^{2\pi} \D \theta~f(\theta)$, where $\theta$ is the polar angle. 

We consider a perturbation $\epsilon$ that causes a splitting of $T_c$ and $T_{\mathrm{TRSB}}$. In the unperturbed case we assume that the two temperatures $T_c^{\epsilon=0} \approx T_{\mathrm{TRSB}}^{\epsilon=0}$ such that the splitting can be meaningfully expanded in the small quantity $y \coloneqq 1 - T_{\mathrm{TRSB}}^{\epsilon}/T_c^{\epsilon}$. We further assume, as is standard, that the magnitudes of the order parameter components scale with the respective temperatures, i.e.,~$\Delta_{0a}^{\epsilon} = \Delta_{0a}^{\epsilon=0} T_c^{\epsilon} / T_c^{\epsilon = 0}$ and $\Delta_{0b}^{\epsilon} = \Delta_{0b}^{\epsilon=0} T_{\mathrm{TRSB}}^{\epsilon} / T_{\mathrm{TRSB}}^{\epsilon = 0}$. Consequentially, we find that $\eta$ to leading order in $y$ is given by
\begin{widetext}
\begin{equation}
\eta = \left( \frac{\Delta_{0b}^{\epsilon=0}}{\Delta_{0a}^{\epsilon=0}}\right)^2  \frac{\langle \lvert f_b(\theta) \rvert^2 \rangle_{\mathrm{FS}} }{\langle \lvert f_a(\theta) \rvert^2 \rangle_{\mathrm{FS}} } \Big[ 1 -  y \frac{7\zeta(3)}{\pi^2}  \left(\frac{\Delta_{0a}^{\epsilon=0} }{2 T_{c}^{\epsilon=0}} \right)^2 \frac{  \langle \lvert f_a(\theta) \rvert^2 \lvert f_b(\theta) \rvert^2 \rangle_{\mathrm{FS} } }{ \langle \lvert f_b(\theta) \rvert^2 \rangle_{\mathrm{FS} } } +\pazocal{O}(y^2) \Big].
\label{eq:EtaExpansion}
\end{equation}
\end{widetext}
To obtain the above expression we made use of the small argument expansion of the dimensionless integral, $\pazocal{I}(z)$, see Appendix~\ref{sec:SimpleTemp}. Notably, we see that the $\pazocal{O}(y)$ correction is negative semi-definite under the above conditions, so the leading effect of, e.g.,~uniaxial in-plane strain ($\epsilon = \varepsilon_{xx}$) is a reduction of the second heat capacity jump. 

From here it is a simple task to calculate the ratios of Fermi surface averages appearing in Eq.~\eqref{eq:EtaExpansion} for, e.g.,~any harmonic of the one-dimensional $D_{4}$ irreps, i.e.,~$f_{A_1}^{(n)}(\theta) = \cos(4n\theta)$, $f_{A_2}^{(n)}(\theta) = \sin([4n+4]\theta)$, $f_{B_1}^{(n)}(\theta) = \cos([4n+2]\theta)$, $f_{B_2}^{(n)}(\theta) = \sin([4n+2]\theta)$. Focusing on the fundamental harmonic $n = 0$ we see that $\langle \lvert f_a^{(0)}(\theta) \rvert^2 \rangle_{\mathrm{FS}} = \frac{1}{k_F},~\frac{1}{2 k_F}$ for $a = A_1,~\lbrace A_2, B_1, B_2 \rbrace$, respectively. Thus, the strongest suppression of the second jump at $y = 0$ occurs for $a = A_1$ (fully gapped $s$-wave) and $b \in \lbrace A_2, B_1, B_2 \rbrace$ ($g_{xy(x^2-y^2)}$, $d_{x^2-y^2}$, and $d_{xy}$, respectively). The largest $\pazocal{O}(y)$ corrections occur when the nodes are maximally shifted, e.g.,~for $a = B_1$ and $b = B_2$. The $\pazocal{O}(y)$ correction is minimal when the nodes are perfectly matching, e.g.,~for $a = b = B_1$. 

\subsection{Demonstration: numerical evaluation}
\label{sec:SingleBand2}
To further illustrate the points made in the preceding sections, we calculate the heat capacity numerically in the case of the simple square lattice. We fix the nearest-neighbor hopping to $t = 100$~meV, the electronic filling to $\expval{n} = 0.6$ and consider the critical temperatures $T_{c1} = 1.5~\textrm{K}$ and $T_{c2} = 1.2~\textrm{K}$. For the order parameters, we consider the cases $s+id_{xy}$ and $d_{x^2-y^2}+ig_{xy(x^2-y^2)}$, and we employ the leading lattice harmonics for the individual components. As for the temperature dependence we use $\Delta_0(T)$ and $X(T)$ as derived in Appendix~\ref{sec:TwoComponent} with Ginzburg--Landau parameters $\alpha_a^0 = \alpha_b^0 = 30~\textrm{meV}^{-1}$ and $\beta_a = \beta_b = 1~\textrm{meV}^{-3}$ fixed for simplicity. Fixing these parameters is done for illustrational purposes and to highlight the role of $\beta_{ab}$.
\begin{figure}[t!bh]
	\centering
	\includegraphics[width=0.95\linewidth]{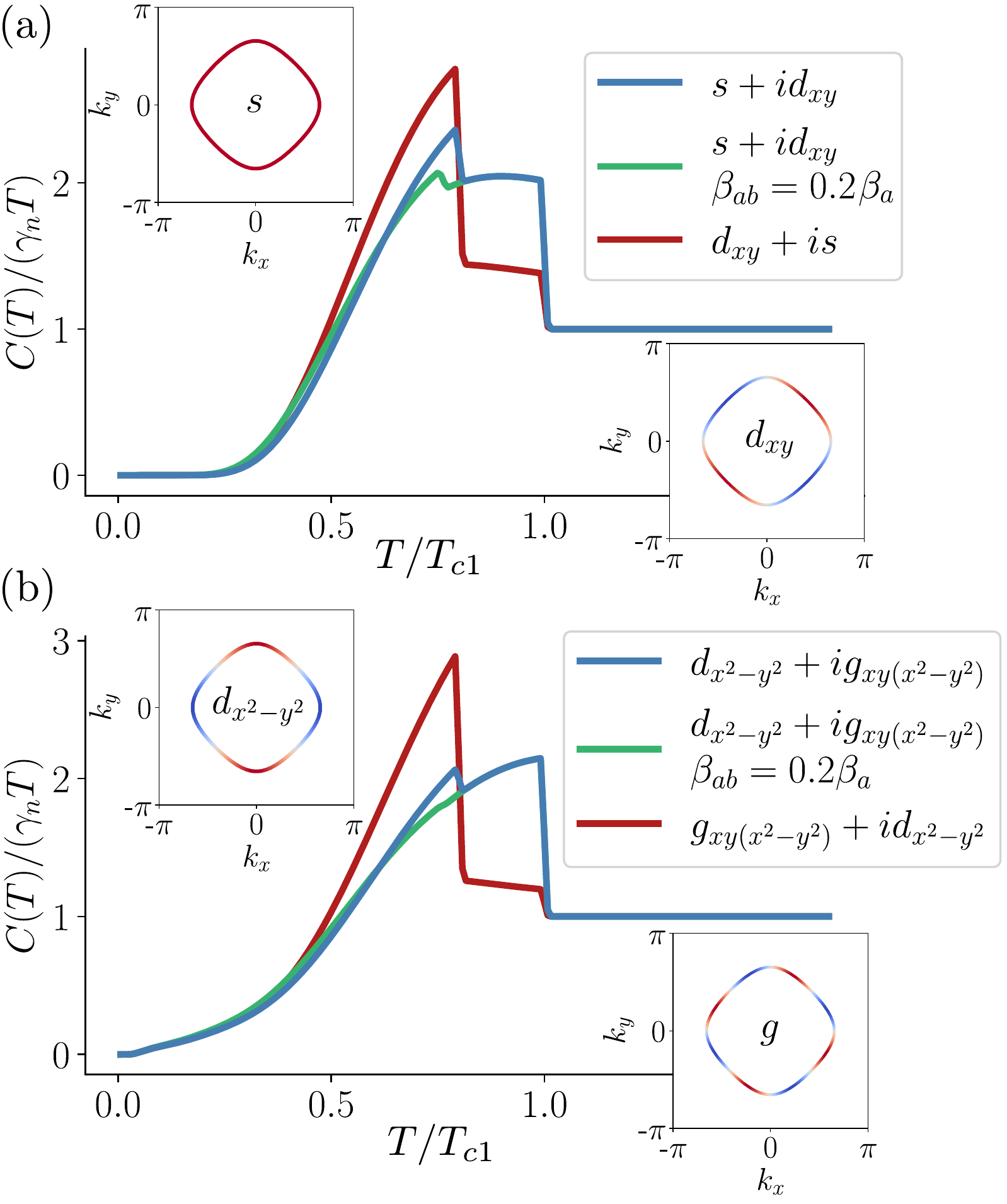}
    \caption{Heat capacity for $s+id_{xy}$ (panel a) and $d_{x^2-y^2} + ig_{xy(x^2-y^2)}$ (panel b) order parameters per temperature per normal-state value in a one-band model. We consider the square lattice with nearest-neighbor hopping $t = 100~\mathrm{meV}$ at filling $\expval{n}=0.6$, with critical temperatures $T_{c1} = 1.5~\textrm{K}$, $T_{c2} = 1.2~\textrm{K}$, and Ginzburg--Landau parameters fixed to $\alpha_a^0 = \alpha_b^0 = 30~\textrm{meV}^{-1}$ and $\beta_a = \beta_b = 1~\textrm{meV}^{-3}$. We take $\beta_{ab}=0$ (blue curves), $\beta_{ab}=0.2\beta_a$ (green curves) and the opposite order of both components with $\beta_{ab}=0$ (red curves), using the fundamental harmonic order parameters shown in the insets.}
	\label{fig:HeatCapacityCircularFS}
\end{figure}

The results clearly confirm the two key points of Sec.~\ref{sec:Parameters}: the second jump is suppressed if the least nodal component onsets first, and turning on $\beta_{ab}$ reduces the second jump further. In the $s+id_{xy}$ case of Fig.~\ref{fig:HeatCapacityCircularFS}(a), the second jump is strongly suppressed with $\beta_{ab} = 0.2 \beta_a$, and it is practically unresolvable under the same conditions for $d_{x^2-y^2}+ig_{xy(x^2-y^2)}$ in Fig.~\ref{fig:HeatCapacityCircularFS}(b). Note that for the order parameter components considered here the low-temperature tail is different for the two scenarios: since $s+id$ is fully gapped, $C(T)/T$ is exponentially suppressed below the full gap, whereas $d+ig$ has symmetry-imposed line nodes along the zone diagonal, making $C(T)/T \sim T$ at low temperatures.

%
\section{Case study: $\mathrm{Sr}_2\mathrm{RuO}_4$}
\label{sec:SRO}
%
Combined heat capacity and muon spin relaxation ($\mathrm{\mu}$SR) measurements have demonstrated that the onset of superconductivity and TRSB split under uniaxial (100) strain in SRO~\cite{GrinekoEA20}. If this splitting is directly linked to the splitting of the critical temperatures of two uniform order parameter components, one generically expects a second heat capacity jump at the lower temperature scale, i.e.,~the onset of TRSB. However, despite intensive efforts from high-resolution heat capacity measurements resolved under strain, no second transition is observed~\cite{LiEA19}. While the absence of a second jump possibly is caused by smearing at zero strain if the two temperatures are finely tuned~\cite{WagnerEA21}, a second jump should in principle be resolvable under finite (100) strain. This conundrum is arguably one of the major outstanding puzzles regarding the superconducting state in Sr$_2$RuO$_4$.

Within recent spin-fluctuation based calculations for superconductivity in Sr$_2$RuO$_4$, it was demonstrated that including longer-range Coulomb interactions in the random phase approximation (RPA) renormalized susceptibility can result in a close competition between $s'$ and $d_{xy}$ solutions to the linearized gap equation, making an $s'+id_{xy}$ order parameter a microscopically viable candidate order parameter for Sr$_2$RuO$_4$~\cite{RomerEA21}. At face value, both accidentally degenerate possibilities $s'+id_{xy}$ and $d_{x^2-y^2}+ig_{xy(x^2-y^2)}$ appear consistent with the breaking of time-reversal symmetry~\cite{Luke98, XiaEA06} with absence of edge currents~\cite{HicksEA10, WangEA22}, as well as nuclear magnetic resonance~\cite{PustogowEA19, ChronisterEA20}, acoustic attenuation~\cite{GhoshEA20, Benhabib2021}, scanning tunneling microscopy~\cite{MadhavenEA19, BhattacharyyaEA21}, and heat conductivity measurements~\cite{HassingerEA17}. However, it should be noted that several outstanding questions remain on how such accidentally degenerate pairing condensates relate to phase sensitive junction experiments~\cite{NelsonEA04, FrancoiseEA06, AnwarEA19}, indirect observations of half-quantum vortices~\cite{JangEA11, YuanEA21}, and field-angle-dependent specific heat~\cite{KittakaEA18}. In addition, accidentally degenerate pairing states face the challenge to explain the missing second specific heat jump, as discussed above. It is this latter question that we further pursue here.

\subsection{Band structure, order parameters, and transition temperatures}
\label{sec:BandsOPTemps}
For the band structure we employ a 2D three-band tight-binding model describing the $t_{2g}$ orbitals of the Ru atoms consistent with angular resolved photoemission spectroscopy (ARPES) measurements~\cite{ZabolotnyyEA13}. Uniaxial (100) strain is modelled by linear distortions, quantified by a dimensionless parameter $p$, of the hopping parameters using the experimentally determined Poisson's ratio~\cite{BarberEA19,RomerEA20}. In Fig.~\ref{fig:BandStructure} we show the Fermi surface obtained from the tight-binding model, with details listed in Appendix~\ref{sec:Bandstructure}. The left panel displays the unstrained band structure ($p = 0$), and the right panel shows the band structure subject to uniaxial (100) strain ($p = -0.075$) close to the van Hove point at which the $\gamma$ band undergoes a Lifshitz transition. 
\begin{figure}[t!bh]
	\centering
	\includegraphics[width=\linewidth]{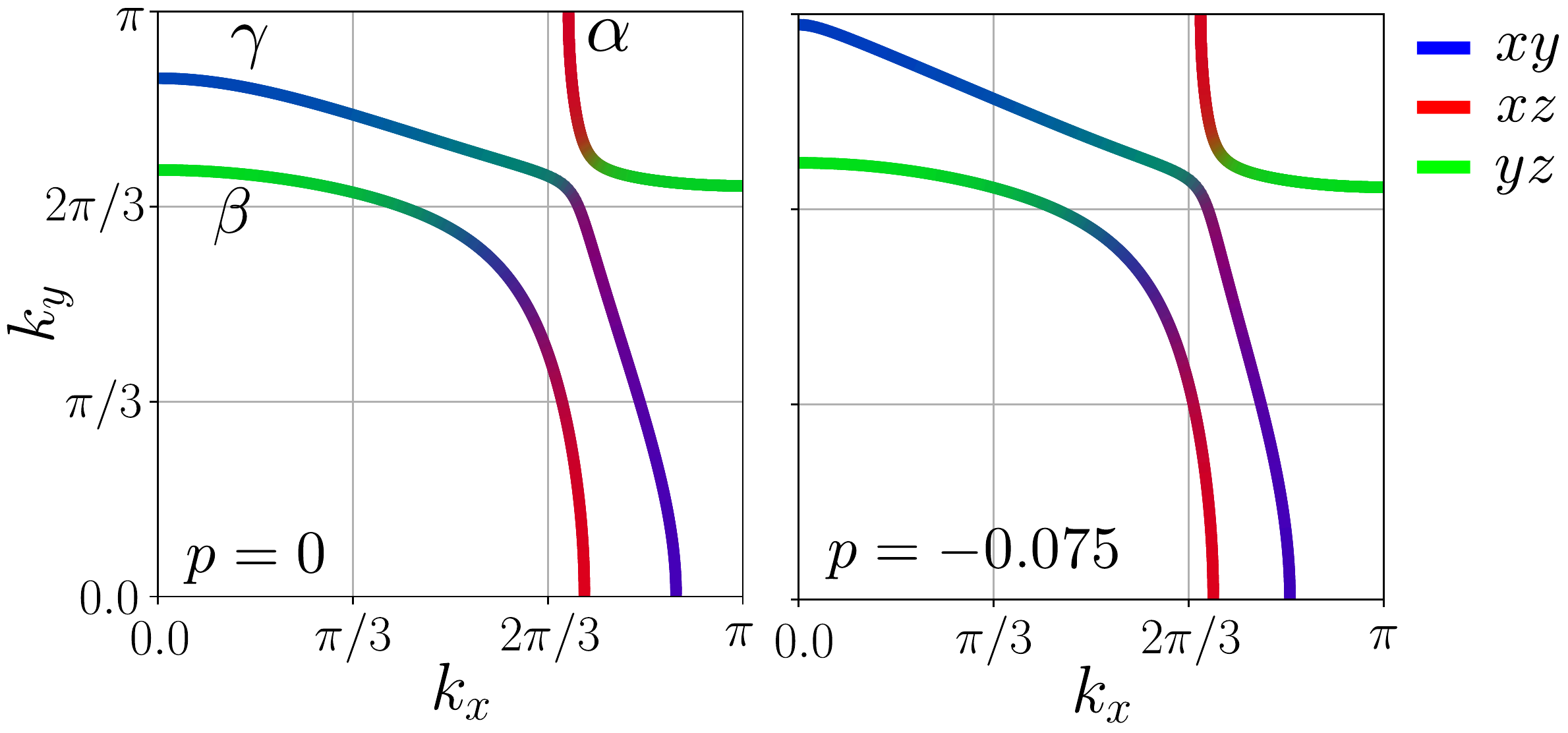}
\caption{Fermi surface from the three-band tight-binding model for Sr$_2$RuO$_4$ described in Appendix~\ref{sec:Bandstructure}. Left panel: unstained case, and right panel: uniaxial (100) strain just below the Lifshitz transition on the $\gamma$ band. The bands have been colour coded by their Ru $t_{2g}$ orbital content.}
	\label{fig:BandStructure}
\end{figure}

As demonstrated in Ref.~\onlinecite{RomerEA21}, for the parameter choice $U = 100$ meV (on-site Hubbard), $J/U = 0.1$ (Hund's coupling), and $V/ U = 0.2$ (nearest-neighbor Hubbard), an $s'+id_{xy}$ solution can be stabilized as a function of $V'/U$, where $V'$ is the next-nearest neighbor interaction. 
\begin{figure}[t!bh]
	\centering
	\includegraphics[width=0.95\linewidth]{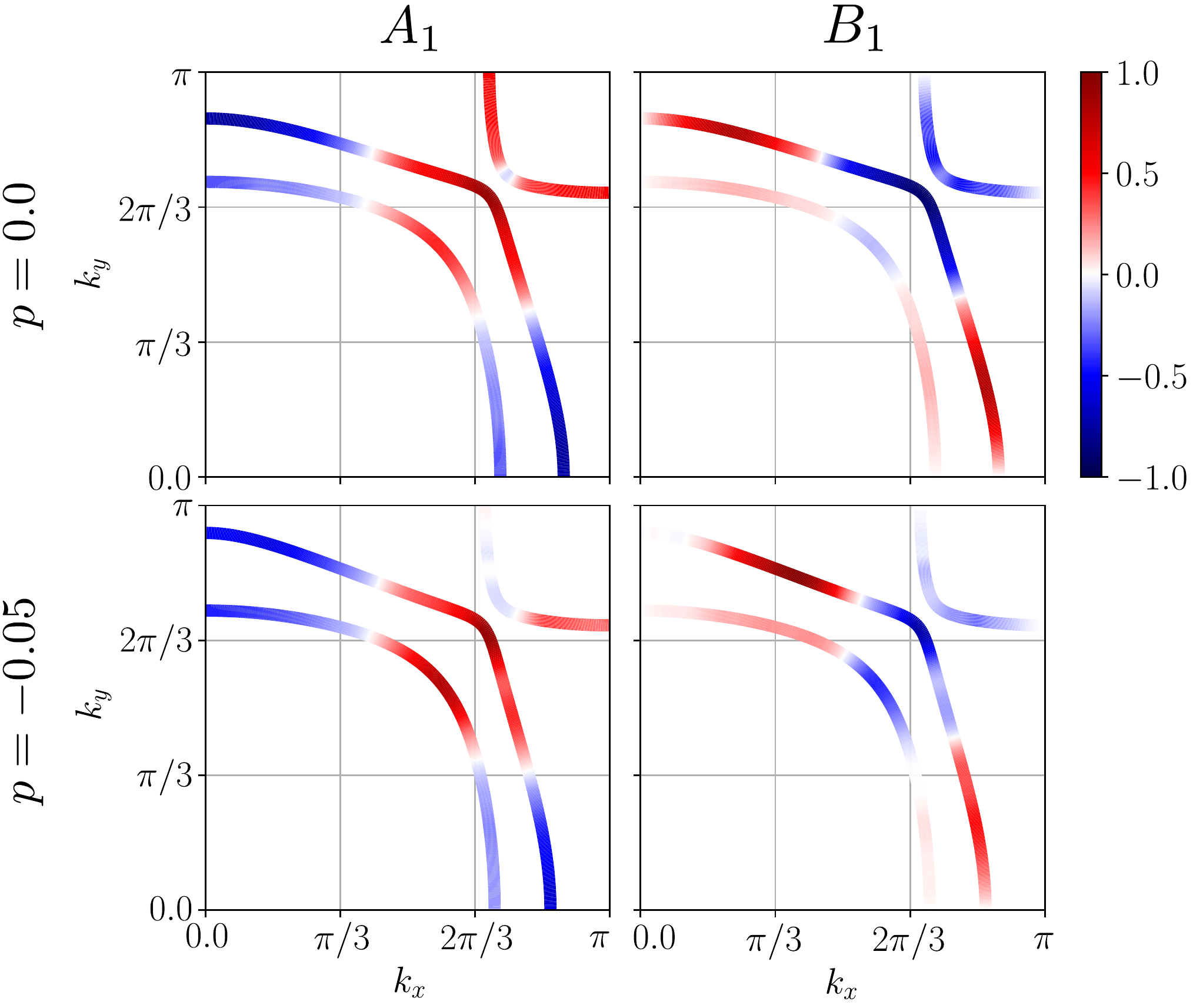}
\caption{Leading even-parity order parameters (labelled $s'$ and $d_{xy}$ at $p=0$) belonging to irreducible representations $A_1$ and $B_1$ of the point group $D_{2}$ at zero and finite (100) uniaxial strain from the RPA calculations of Ref.~\onlinecite{RomerEA21}. The interaction parameters were set to $U = 100$ meV, $J/U = 0.1$, and $V/ U = 0.2$.}
	\label{fig:OrderParameters}
\end{figure}
In Fig.~\ref{fig:OrderParameters} we show the leading and nearly degenerate $s'$ and $d_{xy}$ solutions at zero and finite ($p = -0.05$) uniaxial (100) strain for $V'=0$. The two components share momentum structure on large portions of the Fermi surface, and the combination $s'+id_{xy}$ has a near-node on the $\gamma$ band that persists to finite strains, the position appears well-correlated with where the orbital content changes from being dominated by $d_{xy}$ to $d_{xz/yz}$~\cite{RomerEA21}. Similarly, in Fig.~\ref{fig:OrderParameters2} we show the next-to-leading order parameters ($d_{x^2-y^2}$ and $g_{xy(x^2-y^2)}$) for the same interaction parameters. The next-to-leading orders share, by symmetry, a common node along the zone diagonals $k_x = \pm k_y$ in the unstrained case. This common node is lifted at finite in-plane strain. Supplementary plots of the order parameters are provided in Appendix~\ref{sec:MorePlots}.

\begin{figure}[t!bh]
	\centering
	\includegraphics[width=0.95\linewidth]{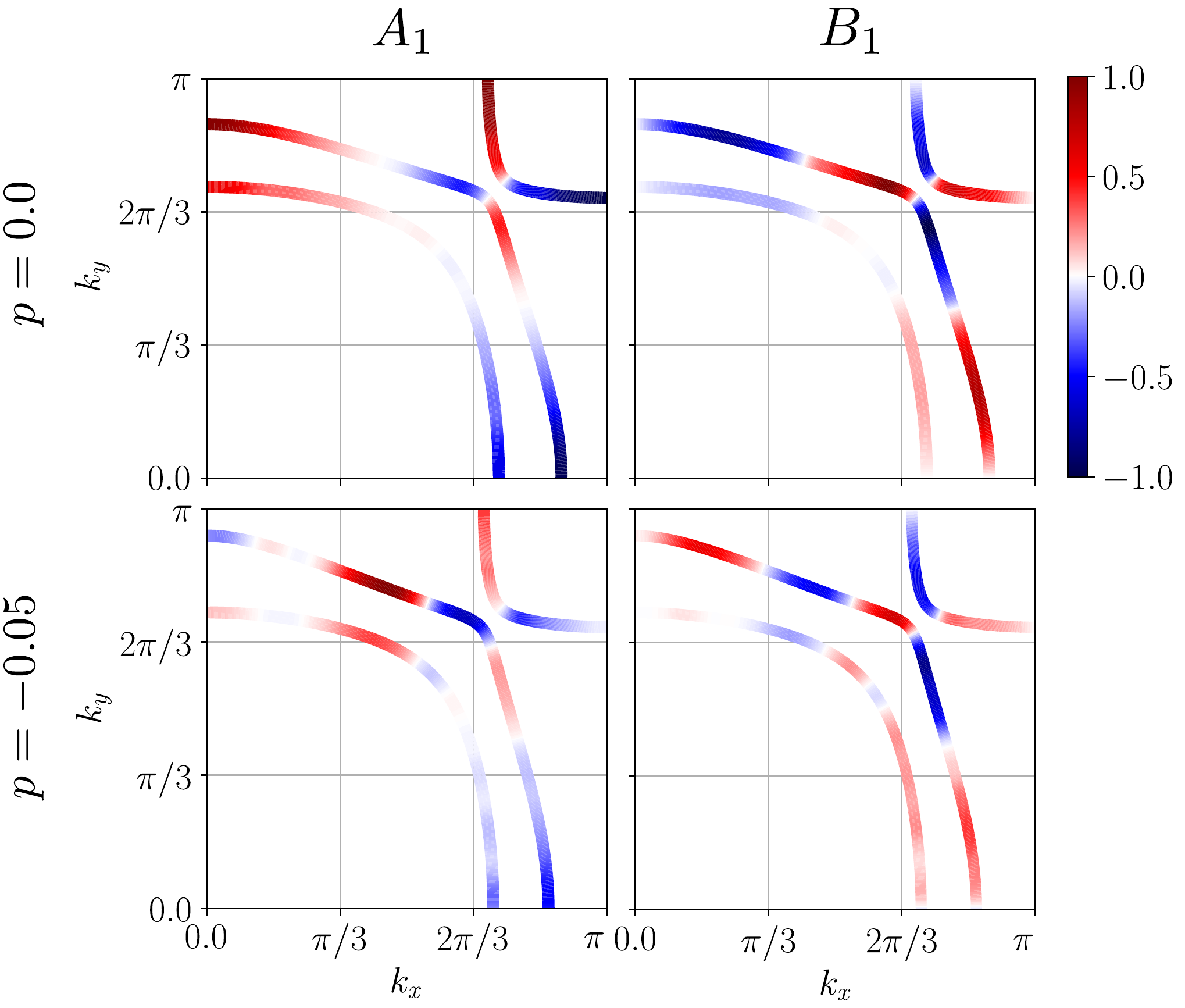}
\caption{Next-to-leading even-parity order parameters (labelled $d_{x^2-y^2}$ and $g_{xy(x^2-y^2)}$ at $p = 0$) belonging to irreducible representations $A_1$ and $B_1$ of the point group $D_{2}$ at zero and finite (100) uniaxial strain from the RPA calculations of Ref.~\onlinecite{RomerEA21} and for the same interaction parameters as in Fig.~\ref{fig:OrderParameters}. }
	\label{fig:OrderParameters2}
\end{figure}

\begin{figure}[t!bh]
	\centering
	\includegraphics[width=0.85\linewidth]{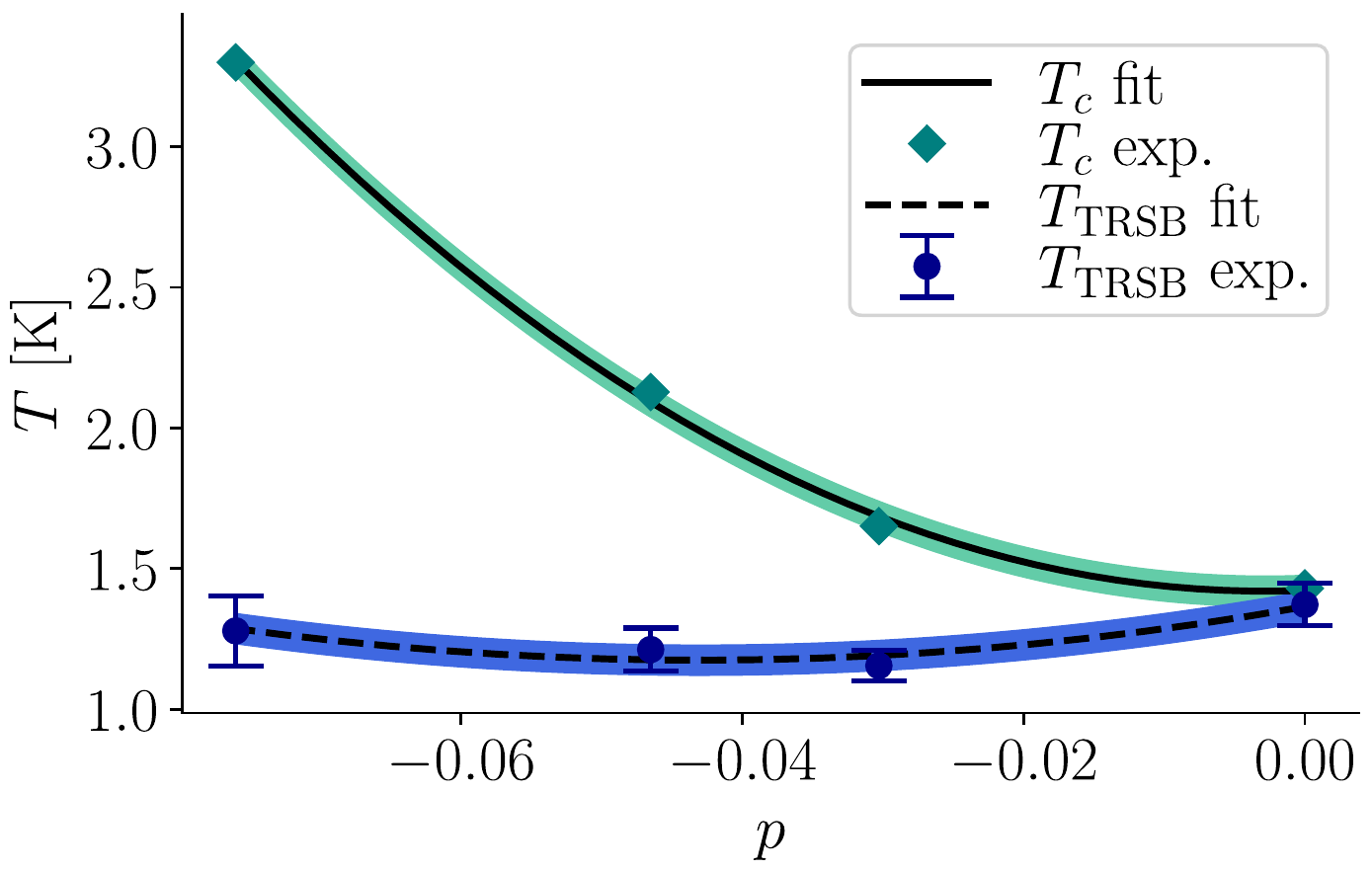}
\caption{Split transition temperatures as a function of the dimensionless uniaxial (100) strain parameter $p$, as measured with combined heat capacity (teal diamonds) and $\mu$SR (blue circles) from Ref.~\cite{GrinekoEA20}. Gray lines are quadratic fits to the data points, used to parameterize $T_c(p)$ and $T_{\mathrm{TRSB}}(p)$ as used in the calculations. The light blue and red coloured bands indicate transition widths of $\sigma = 100$~mK; cf.~the simulated smearing in Fig.~\ref{fig:HeatCapacity}.}
	\label{fig:Temperatures}
\end{figure}

As mentioned above, combined heat capacity and $\mathrm{\mu}$SR measurements have demonstrated that the onset of superconductivity and TRSB split under uniaxial (100) strain in Sr$_2$RuO$_4$~\cite{GrinekoEA20}. Theoretically, this result has been described as a strain-induced splitting of the leading superconducting instabilities~\cite{RomerEA20}. In Fig.~\ref{fig:Temperatures} we show the experimentally obtained transition temperatures, $T_c$ and $T_{\mathrm{TRSB}}$, as obtained from the primary heat capacity jump and the abrupt change in the muon spin relaxation rate, respectively. 
The data points have been fitted with quadratic polynomials after a linear re-scaling of the strain axis such that the experimental van Hove strain of $\varepsilon_{xx} = -0.703$~GPa matches the dimensionless strain parameter $p = -0.076$. 

\subsection{Calculated specific heat transitions}
\label{sec:Results}
\begin{table}[t!bh]
\caption{Calculated ratios of specific heat jumps $\eta$ extracted from the numerical results of Fig.~\ref{fig:HeatCapacity} (c) and (d).}
\begin{center}
\begin{tabular}{p{1.5cm} p{2.8cm} p{2.8cm}}
\toprule
 Strain $p$ & Jump ratio $\eta_{s'+id}$ & Jump ratio $\eta_{d+ig}$ \\ \hline
 $0.0$ & $0.014$ & $0.369$ \\
 $-0.02$ & $0.283$ & $0.453$ \\
 $-0.03$ & $0.198$ & $0.269$ \\
 $-0.04$ & $0.081$ & $0.827$ \\
 $-0.05$ & $0.012$ & $0.938$ \\
 \bottomrule
\end{tabular}
\end{center}
\label{tab:Jumps}
\end{table}
\begin{figure*}[tb]
	\centering
	\includegraphics[width=0.75\linewidth]{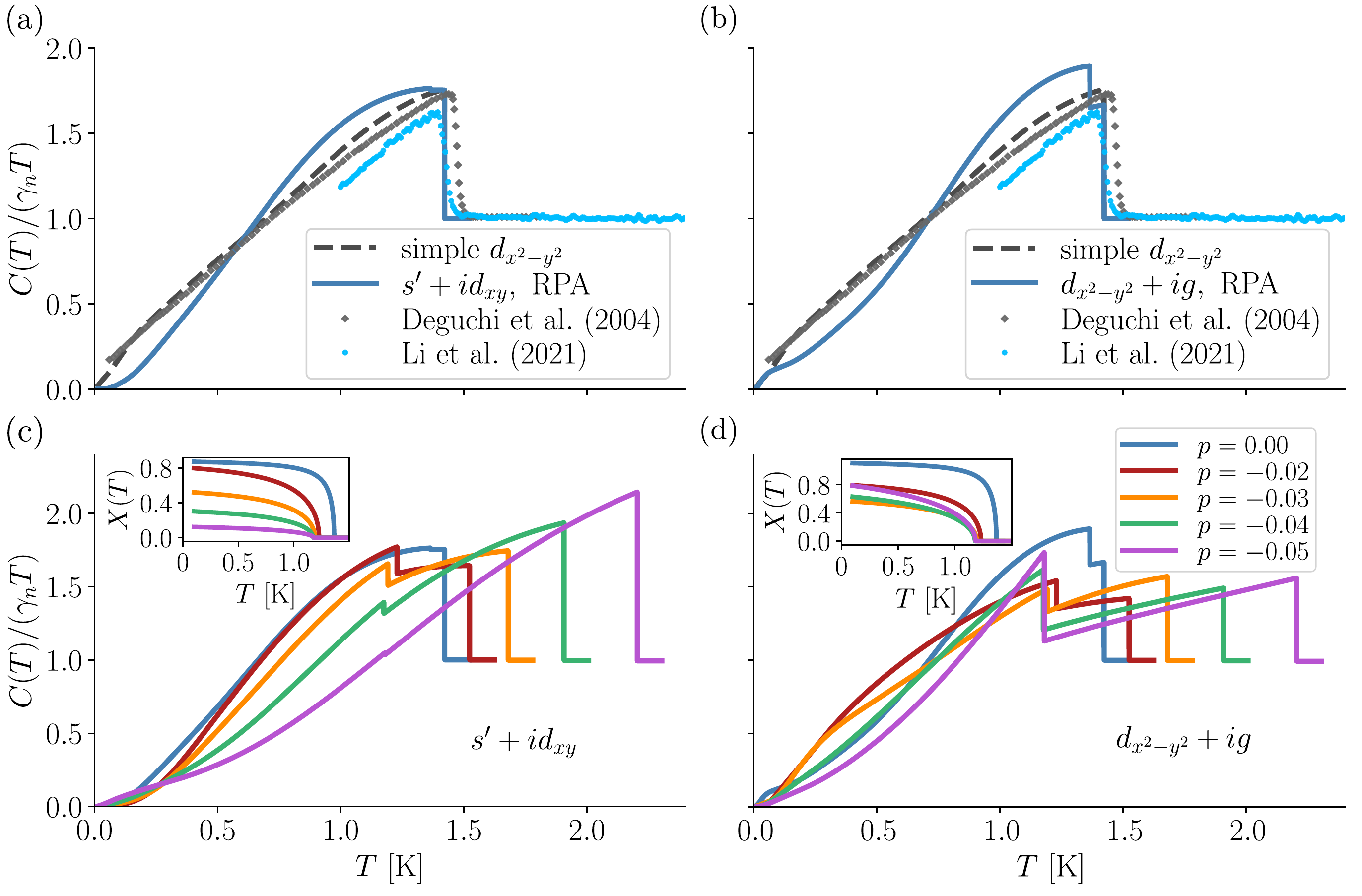} 
\caption{Heat capacity per temperature and normal state value for the (left column) $s'+id_{xy}$ and (right column) $d_{x^2-y^2}+ig_{xy(x^2-y^2)}$ order parameters from Ref.~\onlinecite{RomerEA21}. Panels (a) and (b): the unstrained case. We use (blue lines) the order parameters from Figs.~\ref{fig:OrderParameters} and~\ref{fig:OrderParameters2} with $\Delta_0(T)$ and $X(T)$ (insets) that minimize the free energy of Eq.~\eqref{eq:FreeEnergy}. The experimental data from Refs.~\onlinecite{DeguchiEA04,LiEA19} is plotted alongside. The black dashed line shows the result for a single-component $d_{x^2-y^2}$-wave order parameter, $\Delta_{\mu}(\bo{p}) = \Delta_{\mu} (\cos{p_x}-\cos{p_y})$ with $\Delta_{\alpha}/\Delta_{\gamma} = 0.55$ and $\Delta_{\beta}/\Delta_{\gamma} = 0.45$, for reference. Panels (c) and (d): calculated heat capacity at finite strain with $X(T)$ shown in the insets.}
	\label{fig:HeatCapacity}
\end{figure*}

Equipped with the band structure, order parameters and transition temperatures of the preceding subsection, we numerically compute the Ginzburg--Landau coefficients of Eqs.~\eqref{eq:Alpha}-\eqref{eq:Betaab} and minimize the free energy of Eq.~\eqref{eq:FreeEnergy} with respect to $\Delta_0(T)$ and $X(T)$ (see Appendix~\ref{sec:TwoComponent} for the general expressions). Using the order parameter of Eq.~\eqref{eq:OPansatz}~\footnote{Experimentally, the size of the gap has been inferred from STM measurements to be about $350$ $\mu$eV~\cite{MadhavenEA19}.}, we proceed by evaluating the heat capacity as a function of uniaxial (100) strain. Results are shown in Fig.~\ref{fig:HeatCapacity} for both the $s'+id_{xy}$ (panels a and c) and the $d_{x^2-y^2}+ig_{xy(x^2-y^2)}$ (panels b and d) order parameters. We stress that labelling the order parameters by the $D_{4}$ harmonics, like $d_{x^2-y^2}$, is not strictly accurate notation; by $s'$ we refer to the leading order parameter belonging to irrep $A_1$ of $D_2$, by $d_{xy}$ we refer to the leading order in the $B_1$ channel, by $d_{x^2-y^2}$ we refer to the next-to-leading order in the $A_1$ channel, and by $g_{xy(x^2-y^2)}$ we refer to the next-to-leading order in the $B_1$ channel. For the five considered values of strain, $p = 0.0,-0.02,-0.03,-0.04,-0.05$, the numerical results for the heat capacity jump ratios are summarized in Table~\ref{tab:Jumps}. As seen, the the second discontinuity is generally rather modest for the $s'+id_{xy}$ case, especially at zero strain where the jump is hardly visible in Fig.~\ref{fig:HeatCapacity}(a).

In Fig~\ref{fig:HeatCapacitySmear} we show the simulated effect of smearing of the phase transition, which experimentally could originate from disorder or strain inhomogeneity. The smearing is modelled by averaging over $300$ realizations with the transition temperatures drawn from normal distributions with mean given by the fits $T_c(p)$ and $T_{\mathrm{TRSB}}(p)$ in Fig.~\ref{fig:Temperatures} and standard deviation of $\sigma = 100~\mathrm{mK}$, respectively. In the experiment of Ref.~\onlinecite{LiEA19}, the width of the primary heat capacity jump varies around $\sigma = 30$-$200$~mK. The width of the jump initially increases with strain, before becoming sharper again at the largest values of the strain. This can be understood if the dominant source of the broadening is strain inhomogeneity: At intermediate strain, $T_c$ changes fastest as a function of strain and therefore the effect of strain inhomogeneity is the most pronounced~\cite{LiEA19}. 

The effect of smearing is more easily understood in a toy model for the heat capacity (per temperature)
\begin{equation}
    c(T, T_c) \coloneqq a T\left[ 1-\Theta(T-T_c) \right],
\end{equation}
where $\Theta$ is the Heaviside function and $a$ is the slope of $c$ below $T_c$. If the step locations $T_c$ are drawn from a normal distribution with mean $\bar{T}_c$ and standard deviation $\sigma$, then 
\begin{equation}
    \lim_{N \to \infty} \frac{1}{N}\sum_{\lbrace T_c \rbrace} c(T,T_c) = \frac12 a T \big[1-\mathrm{Erf}\big(\frac{T-\bar{T}_c}{\sqrt{2}\sigma}\big) \big],
\end{equation}
where $\mathrm{Erf}$ is the error function; the width of the smeared transition is straightforwardly controlled by $\sigma$. One may therefore expect a significantly smeared double transition when $T_c - T_{\mathrm{TRSB}} \lesssim \sigma$. In Fig.~\ref{fig:Temperatures} we have indicated widths of $\sigma = 100$~mK as coloured bands around the transition temperatures.

\subsection{Discussion}
\label{sec:Discussion}
Considering the results of the unstrained material in the top panel of Fig.~\ref{fig:HeatCapacity}, neither order parameter candidate, $s'+id_{xy}$ or $d_{x^2-y^2}+ig_{xy(x^2-y^2)}$, provides a perfect match with the temperature dependence of the experimental data of Ref.~\onlinecite{DeguchiEA04}. However, given that we feed in microscopically computed form factors obtained from a specific set of parameters, we do not consider this discrepancy a fundamental concern. As demonstrated in Appendix~\ref{sec:MorePlots}, the match to the experimental data can be improved by, e.g.,~redistributing the band-resolved gap magnitudes compared to the RPA solution.

Noticeably, in the unstrained case the second jump of the $s'+id_{xy}$ order is very tiny, and an order of magnitude smaller than for $d_{x^2-y^2}+ig_{xy(x^2-y^2)}$. Under uniaxial in-plane strain, the second jump is initially found to grow in the case of $s'+id_{xy}$ before decreasing again for strains $p<-0.02$. For $d_{x^2-y^2}+ig_{xy(x^2-y^2)}$ the second jump evolves differently with strain but remains sizeable and outside the threshold for detection reported in Ref.~\onlinecite{LiEA19}. 

The second jump evolving non-monotonically under strain in both order parameter scenarios points to the importance of the $A$-term in Eq.~\eqref{eq:EtaMain} since the simplified case of Eq.~\eqref{eq:EtaExpansion}, in which the $A$-term is absent, predicts a monotonic decrease of the second jump with strain. As noted before, the $A$-term becoming sizeable is linked to a shared momentum structure of the form factors. Accommodating a small second jump under strain thus requires the accidental nodes, which generally shift on the Fermi surface under strain, to shift in a similar fashion for the two form factor components, which is highly constraining and \emph{a priori} implausible. Any failure of this synchronized node-matching under strain can therefore be identified as an effect working against second jump reduction found in Eq.~\eqref{eq:EtaExpansion}. 

In both order parameter scenarios considered here, the $A_1$ and $B_1$ form factors exhibit accidental nodes in the vicinity of $k_x, k_y = \pi/3$ at zero strain. Figures \ref{fig:OrderParameters} and \ref{fig:OrderParameters2} show that for $s'$ and $d_{xy}$ these nodes shift in a rather modest way with strain compared to $d_{x^2-y^2}$ and $g$, the latter two also develop additional nodal regions at the highest strains. It is also worth noting that the symmetry-protected zone-diagonal node in the $d_{x^2-y^2}+ig_{xy(x^2-y^2)}$ scenario is lifted for finite strain, which can be seen in the low-temperature tail in Fig.~\ref{fig:HeatCapacity}(d). We attribute the dissimilar behaviour of the second jump under strain of $s'+id_{xy}$ vs. $d_{x^2-y^2}+ig_{xy(x^2-y^2)}$ to this different strain-induced node shifting.

\begin{figure}[t!bh]
	\centering
	\includegraphics[width=0.90\linewidth]{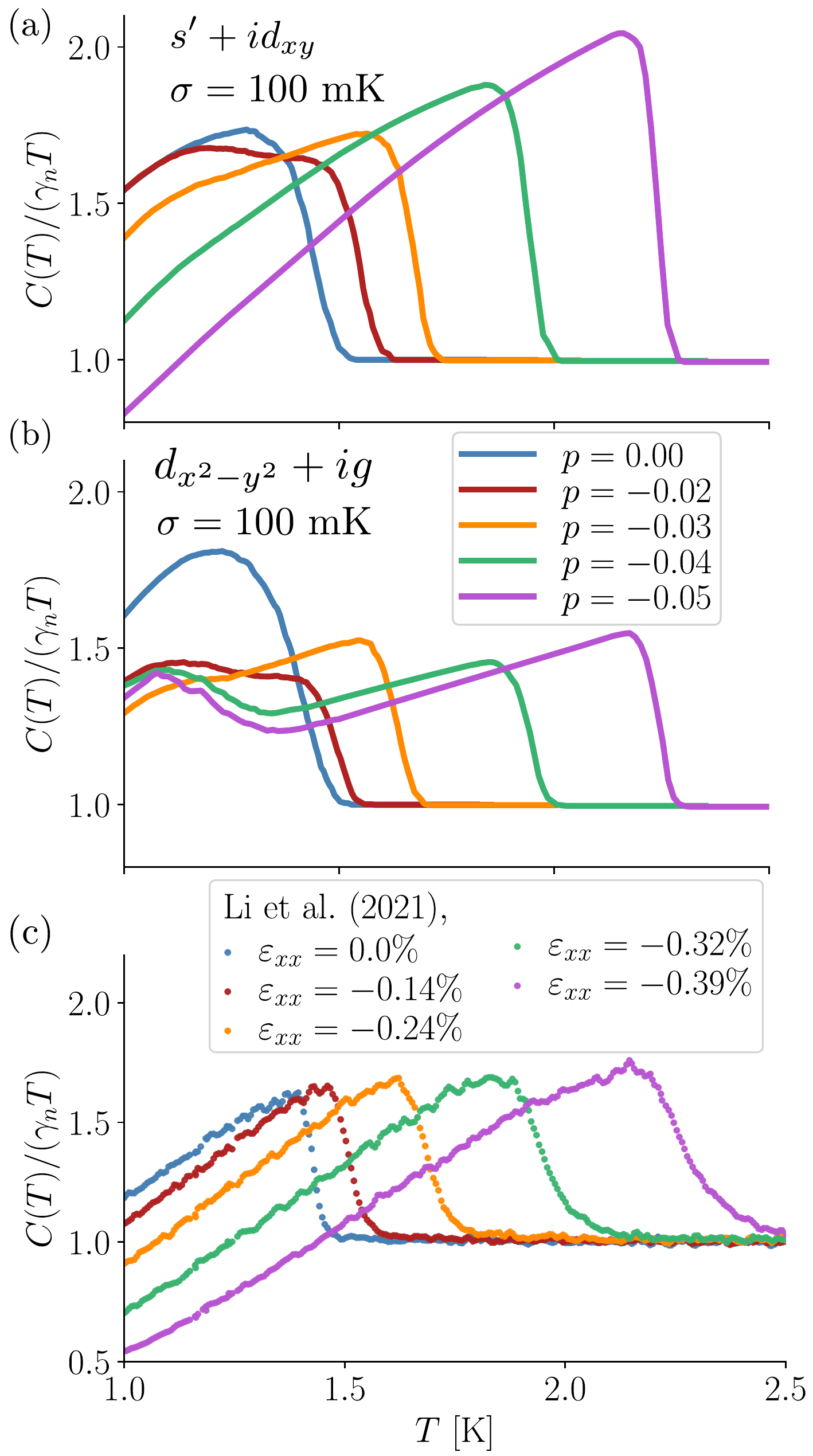}
\caption{Simulated effect of smearing in the (a) $s'+id_{xy}$ and (b) $d_{x^2-y^2}+ig_{xy(x^2-y^2)}$ order parameter scenario for the same strain values as considered in Fig.~\ref{fig:HeatCapacity} (c) and (d). Smearing is here modelled by averaging over realizations where we draw transition temperatures from normal distributions with means $T_c(p)$ and $T_{\mathrm{TRSB}}(p)$ from Fig.~\ref{fig:Temperatures} and standard deviation $\sigma=100~\mathrm{mK}$. For each realization we calculate the Ginzburg--Landau coefficients of Eqs.~\eqref{eq:Alpha}-\eqref{eq:Betaab} and obtain the heat capacity from Eq.~\eqref{eq:SpecificHeatNormalized} with the order parameter of Eq.~\eqref{eq:OPansatz}. Panel (c): The experimental data from Ref.~\onlinecite{LiEA19}, providing an indirect measurement of the heat capacity.}
	\label{fig:HeatCapacitySmear}
\end{figure}

When smearing is added in Fig.~\ref{fig:HeatCapacitySmear}, it is seen that a smearing width of $\sigma = 100~\mathrm{mK}$ is enough to bring the second jump into the undetectable regime for the $s'+id_{xy}$ case; a case with $50~\mathrm{mK}$ smearing is shown Appendix.~\ref{sec:MorePlots}. In contrast, the second jump remains clearly resolvable for $d_{x^2-y^2}+ig_{xy(x^2-y^2)}$. When comparing to the experimental data of Ref.~\onlinecite{LiEA19}, shown in Fig.~\ref{fig:HeatCapacitySmear}(c) for completeness, we conclude that $s'+id_{xy}$ is a more likely order parameter candidate than $d_{x^2-y^2}+ig_{xy(x^2-y^2)}$ within the premises of our calculation. We also note that the results change drastically if one instead considers the reversed combination of orders, e.g.,~$d_{xy} + is'$ as opposed to $s'+id_{xy}$. In Appendix~\ref{sec:MorePlots} we show results for the $d_{xy} + is'$ case, which results in $\eta \geq 0.26$. Thus, as noted in Sec.~\ref{sec:Parameters}, the order in which the order parameters switch on has a large impact on the heat capacity jumps. The order parameter combination $s'+id_{xy}$ where the $s'$ order condenses at the higher temperature has a small second jump, whereas the $d_{xy}+is'$ combination has a large second jump and can therefore be ruled out.

Among the aforementioned order parameter candidates recently proposed for Sr$_2$RuO$_4$ we have so far not addressed the $d_{xz}+id_{yz}$ state ($E_g$ irrep of $D_{4h}$)~\cite{ZuticEA05, SuhEA19}. Although this order parameter can not be accessed with a two-dimensional model for the band structure, we note that previous microscopic three-dimensional models have found this state unfavourable~\cite{RoisingEA19, JerzembeckEA21, RomerEA22}. Yet, based on the principles regarding the size of the second heat capacity jump in Sec.~\ref{sec:HeatCapacityMain}, we can infer general expectations for split transitions in the case of two-dimensional irreps, even though the quantitative result comes down to details of the specific order parameters considered. Note that if the two components are degenerate by symmetry, the perturbation $\epsilon$ that causes their splitting must break the crystal symmetries, like uniaxial $(100)$ strain in tetragonal systems. Besides the symmetry-imposed horizontal line node at $k_z = 0$ of $d_{xz}+id_{yz}$, the vertical symmetry-imposed nodes of the two components, i.e.,~at $k_x = 0$ and $k_y = 0$, are maximally shifted which from the principles of Sec.~\ref{sec:Parameters} is expected to give a sizeable second jump. Furthermore, since the sizes of the two components are equal in the unperturbed case by symmetry ($X = 1$), the two transitions are naively expected to be associated with equally-sized heat capacity jumps, at least close to the unperturbed case (cf.~supplemental Fig.~S9 of Ref.~\onlinecite{LiEA19}). Thus, based on these considerations alone, the $d_{xz}+id_{yz}$ scenario for Sr$_2$RuO$_4$ appears less likely than the accidentally degenerate cases of $s'+id_{xy}$ and $d_{x^2-y^2}+ig_{xy(x^2-y^2)}$. 

Contrasting the focus on exotic multi-component orders relying on accidental degeneracies, the simpler explanation of the specific heat data of Sr$_2$RuO$_4$ in terms of a single $d_{x^2-y^2}$ order parameter should not be dismissed (cf.~the gray dashed line in Fig.~\ref{fig:HeatCapacity}). As discussed above, however, this possibility raises the immediate question of how to explain, e.g., the ultrasound experiments and the evidence of TRSB~\cite{GhoshEA20, Benhabib2021, XiaEA06, GrinenkoEA21}. Regarding the latter, since Sr$_2$RuO$_4$ resides in the vicinity of magnetic instabilities~\cite{OrtmannEA13} perhaps TRSB is related to glassy pinned magnetism or coupling between magnetic and superconducting orders~\cite{Andersen007,Gastiasoro16}. Another possibility is that there is a second inhomogeneous order parameter component that is induced by sample dislocations, which is also expected to result in a small second jump~\cite{WillaEA21}.

%
\section{Other time-reversal symmetry broken superconductors}
\label{sec:OtherMaterials}
%
In the heavy fermion compound UPt$_3$ there is convincing experimental evidence of a two-component order parameter deep in the superconducting phase. Two distinct transition temperatures are revealed by two jumps visible in the ultrasound velocity as a function of temperature~\cite{BrulsEA90}. Furthermore, TRSB is observed at temperatures below the second transition. The Kerr effect has been observed~\cite{SchemmEA14}, indicating TRSB below $T_\textrm{TRSB}\approx0.46$K, while the critical temperature as measured by the drop in the resistance is $T_c\approx0.55$K. In addition, TRSB is also seen in a $\mu$SR signal~\cite{LukeEA93} (although we should caution that an independent $\mu$SR experiment failed to see any signature of TRSB~\cite{DalmasEA95}). A likely order parameter candidate is $f_{z(x^2-y^2)}+if_{xyz}$~\cite{JoyntEA02, Kallin16} (where the degeneracy of the two-dimensional irrep is split by magnetism). In UPt$_3$ a very clear second heat capacity jump is observed~\cite{FisherEA89}. The Fermi surface of UPt$_3$ is complicated and contains five strongly three-dimensional sheets when the $5f$ uranium electrons are included~\cite{McMullanEA08}. For the $f_{z(x^2-y^2)}+if_{xyz}$ scenario, modulo all the details of the multiband fermiology and possible accidental nodes, the vertical symmetry-imposed nodes are maximally shifted when comparing the two components, which, according to Sec.~\ref{sec:Parameters} suggests that one can expect equally-sized heat capacity jumps. On the phenomenological level this agrees with what is observed experimentally~\cite{FisherEA89, TrappmannEA91}.

UTe$_2$ has attracted significant attention recently as an unconventional superconductor. The Kerr effect indicating TRSB has been observed~\cite{HayesEA21}. However, a second heat capacity jump is only observed under hydrostatic pressure~\cite{BraithwaiteEA19, ThomasEA20} while no second jump is seen under applied stress~\cite{GirodEA22}. It is hypothesized that the presence or absence of the second jump may be affected by sample quality~\cite{RosaEA22}. UTe$_2$ has an orthorhombic structure ($D_{2h}$) and its Fermi surface, although not fully characterized, is understood to be fully three-dimensional~\cite{AokiEA22}. A large upper critical field indicates that UTe$_2$ is a spin-triplet superconductor and different order parameter combinations that have been proposed include $A_{1u}+iB_{3u}$~\cite{UTe2_OP1}, $B_{1u}+iB_{2u}$ \cite{UTe2_OP2}, and $B_{2u}+iB_{3u}$~\cite{HayesEA21}. In all three cases, the order parameters have different momentum space structure and quasiparticles are gapped out at the second transition, implying that all three order parameter combinations are consistent with a large second heat capacity jump.

Another material relevant for multi-component superconductivity is UBe$_{13}$. In this material, a second heat capacity jump is observed under doping of UBe$_{13}$ to U$_{1-x}$Th$_x$Be$_{13}$~\cite{OttEA85}. $\mu$SR experiments observe TRSB consistent with either TRSB superconductivity or with antiferromagnetism coexisting with superconductivity~\cite{HeffnerEA90}. Note that the $\mu$SR transition coincides with the second heat capacity jump, implying they have a common origin. In more recent experiments~\cite{ShimizuUBe13} two heat capacity jumps were clearly seen as well, and furthermore the low-temperature phase was shown to be fully gapped. This led the authors to propose two different candidates for the two orders. The symmetry group is the cubic space group $O_h$ which has a two-dimensional irrep $E_u$ with the basis functions $\bo{d}_1(k)=\sqrt{3}(k_x\hat x-k_y\hat y)$ and $\bo{d}_2(k)=-k_x\hat x-k_y\hat y+2k_z\hat{z}$. Assuming the degeneracy between the two orders to be lifted with $\bo{d}_1$ switching on at a higher temperature, we expect a large heat capacity jump since $\boldsymbol{d}_1$ has a line node, whereas $\bo{d}_1+i\bo{d}_2$ only has a point node, leading to a significant number of quasiparticles being gapped out at the second transition. The second scenario proposed by the authors considers an accidental degeneracy of $A_{1u}$ and $A_{2u}$, $A_{1u}$ is already fully gapped and therefore based on our formalism we would expect a smaller second jump for the candidate $A_{1u}+iA_{2u}$, making this a less likely candidate. 

Finally, TRSB has been also observed in PrOs$_4$Sb$_{12}$, which has a tetrahedral crystal symmetry $T_h$~\cite{GoryoPrOsSb}, via $\mu$SR experiments~\cite{AokiEA03, ShuEA11}. A clear second heat capacity jump is observed~\cite{MapleEA03,MeassonEA04}. Thermal conductivity measurements indicate the presence of nodes or near-nodes in the high temperature phase and this has lead to the proposal of an $s+id_{z^2-x^2}$ order parameter~\cite{GoryoPrOsSb}, where the $s$-wave order is highly anisotropic and features near-nodes. This could then also lead to a large second heat capacity jump and indeed theoretical calculations suggest that the second jump would be of the same size as the first one for this order parameter~\cite{GoryoPrOsSb}. 

%
\section{Conclusions}
\label{sec:Conclusions}
We have provided a general theoretical framework for the specific heat transitions in multi-component time-reversal symmetry broken superconductors. This allowed a thorough discussion of the amplitudes of the specific heat jumps at the two transition temperatures. Specifically, one expects comparable specific heat discontinuities for most two-dimensional irreps due to the maximally misaligned nodes and the symmetry-imposed equality of order parameter component magnitudes. On the other hand, a significantly reduced second jump may be expected for TRSB superconductors that have competing order parameter components with similar momentum structure, and where the least nodal component onsets first. We applied the theoretical setup to perform a comprehensive discussion of the low-temperature electronic specific heat relevant for Sr$_2$RuO$_4$. While the specific heat data alone for this material are most straightforwardly explained by a nodal single-component superconductor, several other experimental probes point to TRSB multi-component superconductivity~\cite{GhoshEA20, Benhabib2021,XiaEA06,GrinenkoEA21}. We have pursued the accidentally degenerate possibilities $s'+id_{xy}$ and $d_{x^2-y^2}+ig_{xy(x^2-y^2)}$ with gap structures obtained from microscopic RPA calculations, and compared their specific heat behaviour with experimental data in both the absence and presence of uniaxial strain. Our results yield a better agreement between experimental data and   $s'+id_{xy}$ order, as compared to $d_{x^2-y^2}+ig_{xy(x^2-y^2)}$ superconductivity. Further, we conclude from these calculations that the currently-available experimental specific heat data are consistent with  $s'+id_{xy}$ superconductivity in the unstrained case, and also with the uniaxially strained data when considering the role of strain inhomogeneity.

\begin{acknowledgments}
HSR acknowledges helpful conversations with P.~J.~Hirschfeld, A.~Ramires, and M.~H.~Christensen. We thank Olivier Gingras and Ilya Eremin for insightful comments on the manuscript. HSR was supported by a research grant (40509) from VILLUM FONDEN. GW acknowledges funding from the European Research Council (ERC) under the European Union’s Horizon 2020 research and innovation program (ERC-StG-Neupert-757867-PARATOP). ATR acknowledges support from the Danish Agency for Higher Education and Science. ATR and BMA acknowledge support from the Independent Research Fund Denmark grant number 8021-00047B. MR acknowledges support from the Novo Nordisk Foundation grant NNF20OC0060019.
\end{acknowledgments}

\bibliography{SRO}

\onecolumngrid
\begin{appendix}

%
\section{Heat capacity of superconductors}
\label{sec:HeatCapacity}
%
The specific heat for a multiband superconductor follows from the expression for the entropy of the fermion gas~\cite{Tinkham75},
\begin{equation}
S = -2k_B \sum_{\bo{k}, \mu} \Big[ n_F[E_{\mu}(\bo{k})] \ln n_F[E_{\mu}(\bo{k})] + \big(1-n_F[E_{\mu}(\bo{k})] \big) \ln\big( 1 - n_F[E_{\mu}(\bo{k})] \big) \Big].
\label{eq:Entropy}
\end{equation}
Here, $E_{\mu}(\bo{k}) = \sqrt{ \xi_{\mu}^2 + \lvert \Delta_{\mu}(T, \bo{k}) \rvert^2}$, and $n_F(E) = (1+\exp(\beta E))^{-1}$ is the Fermi function. 

\subsection{Single-component order parameter}
\label{sec:SingleCompOP}
We first assume a single-component order parameter with temperature dependence $\Delta(T)$,
\begin{equation}
\Delta_{\mu}(T, \bo{k}) = \Delta(T) f_{\mu}(\bo{k}),
\label{eq:GapFactorizationAnsatz}
\end{equation}
where $f_{\mu}(\bo{k})$ is the order parameter form factor of band $\mu$ normalized such that $\max_{\bo{k},\mu} \lvert f_{\mu}(\bo{k}) \rvert = 1$. The specific heat follows from 
\begin{equation}
\begin{aligned}
C(T) &= T \f{\partial S}{\partial T} = 2 \sum_{\bo{k}, \mu} E_{\mu}(\bo{k}) \f{\D n_F(E_{\mu}(\bo{k}))}{\D T} \\
&= \f{2}{k_B T^2} \int_{-\infty}^{\infty} \D \xi_{\mu} \hspace{1mm} \Big\langle \f{\xi_{\mu}^2 + \Delta(T)^2 \lvert f_{\mu}(\bo{k}) \rvert^2 - \f{T}{2} \f{\partial \Delta(T)^2}{\partial T}  \lvert f_{\mu}(\bo{k}) \rvert^2}{ 4 \cosh^2(\f{E_{\mu}(\bo{k})}{2k_B T})} \Big\rangle_{\mathrm{FS}},
\end{aligned}
\label{eq:Specificheat}
\end{equation}
where the Fermi surface average is evaluated as
\begin{equation}
\langle A \rangle_{\mathrm{FS}} \coloneqq \sum_{\mu} \int_{S_{\mu}} \frac{\D \bo{k}}{(2\pi)^d} \frac{A}{v_{\mu}(\bo{k})},
\label{eq:NewAverage}
\end{equation}
where $v_{\mu}(\bo{k}) = \lvert \nabla \xi_{\mu}(\bo{k}) \rvert$ is the Fermi velocity and $\rho_{\mu} =\frac{1}{(2\pi)^d} \int_{S_{\mu}} \D \bo{k} / v_{\mu}(\bo{k})$ is the density of states of band $\mu$.

By evaluating Eq.~\eqref{eq:Specificheat} in the normal state, for which $\Delta = 0$, the specific heat per temperature, i.e.,~the Sommerfeld coefficient $\gamma_n$, becomes
\begin{equation}
\gamma_n \coloneqq \f{C_n(T)}{ T} = \f{2}{k_B T^3} \sum_{\mu} \rho_{\mu}  \int_{-\infty}^{\infty} \D \xi \f{\xi^2}{4 \cosh^2(\f{\xi}{2k_B T})} = 4 k_B^2 \zeta(2) \sum_{\mu} \rho_{\mu},
\label{eq:NormalStateGamman}
\end{equation}
with $\zeta(2) = \pi^2/6$. This recovers the multiband version of the well-known (electronic) linear contribution to the specific heat of a metal~\cite{Tinkham75}. It is convenient to normalise Eq.~\eqref{eq:Specificheat} by the normal-state value:
\begin{equation}
\frac{C(T)}{T \gamma_n} = \frac{3}{4\pi^2(k_B T)^3 \sum_{b} \rho_{b} } \int_{- \infty}^{\infty} \D \xi \hspace{1mm} \Big\langle \f{\xi^2 + \Delta(T)^2 \lvert f_{\mu}(\bo{k}) \rvert^2 - \f{T}{2} \f{\partial \Delta(T)^2}{\partial T}  \lvert f_{\mu}(\bo{k}) \rvert^2}{\cosh^2(\f{E_{\mu}(\bo{k})}{2k_B T})} \Big\rangle_{\mathrm{FS}}.
\label{eq:SpecificHeatNormalized}
\end{equation}
\subsection{Two-component order parameter with simple temperature dependence}
\label{sec:SimpleTemp}
Following Ref.~\onlinecite{KivelsonEA20} we assume a TRSB two-component order parameter with the particular BCS-like temperature dependence
\begin{equation}
\begin{aligned}
\Delta(\bo{p}) &= \Delta_{0, a} \sqrt{ 1- \frac{T}{T_{c}} } f_{a}(\bo{p})+ i \Delta_{0, b} \sqrt{ 1 - \frac{T}{T_{\text{TRSB}}} } f_{b }(\bo{p}),
\end{aligned}
\label{eq:KivelsonAnsatz}
\end{equation}
where $a$ and $b$ are irrep.~labels. We shall use this ansatz in Eq.~\eqref{eq:Specificheat} and evaluate the heat capacity jump across $T_{c}$ and $T_{\mathrm{TRSB}}$. At the onset of superconductivity we get
\begin{equation}
\begin{aligned}
\frac{\Delta C_{\mathrm{SC}}}{T_c} &\coloneqq \frac{C(T \to T_c^-) - C(T \to T_c^+)}{T_c} \\
&= \frac{1}{T_c} \left( \frac{\Delta_{0,a}}{k_B T_c} \right)^2 \langle \lvert f_a(\bo{p}) \rvert^2 \rangle_{\mathrm{FS}} \int_{-\infty}^{\infty} \frac{\D \xi}{4\cosh^2\left(\frac{\xi}{2k_BT_c}\right)} \\
&= k_B \left( \frac{\Delta_{0,a}}{k_B T_c} \right)^2 \langle \lvert f_a(\bo{p}) \rvert^2 \rangle_{\mathrm{FS}},
\end{aligned}
\label{eq:SpecificHeatJumpTc}
\end{equation}
where the only contribution we pick up comes from the third term in the numerator of Eq.~\eqref{eq:Specificheat}, due to its discontinuity at $T_c$. Repeating the calculation at the onset of the second order parameter component we now get
\begin{equation}
\begin{aligned}
\frac{\Delta C_{\mathrm{TRSB}}}{T_{\mathrm{TRSB}}} &\coloneqq \frac{C(T \to T_{\mathrm{TRSB}}^-) - C(T \to T_{\mathrm{TRSB}}^+)}{T_{\mathrm{TRSB}}} \\
&= k_B \left( \frac{\Delta_{0,b}}{k_B T_{\mathrm{TRSB}}} \right)^2 \Big\langle \lvert f_b(\bo{p}) \rvert^2 \int_{0}^{\infty} \frac{\D u}{\cosh^2\left(\sqrt{u^2+z(\bo{p})^2}\right)} \Big\rangle_{\mathrm{FS}},
\end{aligned}
\label{eq:SpecificHeatJumpTTRSB}
\end{equation}
where $z(\bo{p}) = \frac{ \Delta_{0,a}}{2k_B T_{\mathrm{TRSB}}} \sqrt{1-\frac{T_{\mathrm{TRSB}}}{T_c}} f_a(\bo{p})$ was introduced. Combining these two expressions we get the ratio of the heat capacity jumps
\begin{equation}
\eta \coloneqq \frac{\Delta C_{\mathrm{TRSB}} / T_{\mathrm{TRSB}} }{\Delta C_{\mathrm{SC}} / T_c } = \left( \frac{\Delta_{0,b}}{\Delta_{0,a}} \frac{T_c}{T_{\mathrm{TRSB}}} \right)^2 \frac{\langle \lvert f_b(\bo{p}) \rvert^2 \pazocal{I}[z(\bo{p})] \rangle_{\mathrm{FS}}}{\langle \lvert f_a(\bo{p}) \rvert^2 \rangle_{\mathrm{FS}}},
\label{eq:EtaKivelson}
\end{equation}
where 
\begin{equation}
\pazocal{I}(z) \coloneqq \int_{0}^{\infty} \D u~\cosh^{-2}\left(\sqrt{u^2+z^2}\right).
\label{eq:IntegralI}
\end{equation} 
We note that this function for small $z$ can be approximated as $\pazocal{I}(z) = 1 - \frac{7\zeta(3)}{\pi^2} z^2 + \pazocal{O}(z^4) \approx \exp\left(-\frac{7\zeta(3)}{\pi^2} z^2\right)$.

\subsection{Two-component order parameter with general temperature dependence}
\label{sec:BeyondKivelson}
We go beyond the square root ansatz in Eq.~\eqref{eq:KivelsonAnsatz} by assuming
\begin{equation}
\Delta(T, \bo{p}) = \Delta_0(T) \left[ f_{a}(\bo{p}) + i X(T) f_{b}(\bo{p}) \right].
\label{eq:BeyondKivelsonAnsatz}
\end{equation}
Above we have left band indices on $f_{a/b}(\bo{p})$ implicit. Crucially, this ansatz allows for the two components to interact in the sense that $\Delta_0(T)$ can, and generally will, be non-analytic at $T = T_{\mathrm{TRSB}}$ where $X$ onsets. This also results in two terms being picked up in the numerator of $\eta$, coming from 
\begin{equation}
A \coloneqq \left[ \frac{\partial \Delta_0(T)^2}{\partial T}\Big\rvert_{T \to T_{\mathrm{TRSB}}^-} - \frac{\partial \Delta_0(T)^2}{\partial T}\Big\rvert_{T \to T_{\mathrm{TRSB}}^+} \right] \big/ D
\label{eq:delta1} 
\end{equation}
and 
\begin{equation}
B \coloneqq \left[ \frac{\partial (\Delta_0(T)X(T))^2}{\partial T}\Big\rvert_{T \to T_{\mathrm{TRSB}}^-} -\frac{\partial (\Delta_0(T)X(T))^2}{\partial T}\Big\rvert_{T \to T_{\mathrm{TRSB}}^+}\right] \big/ D,
\label{eq:delta2} 
\end{equation}
with $D \coloneqq \frac{\partial \Delta_0(T)^2}{\partial T}\big\rvert_{T \to T_c^-} - \frac{\partial \Delta_0(T)^2}{\partial T}\big\rvert_{T \to T_c^+}$, respectively. Repeating the calculation of Sec.~\ref{sec:SimpleTemp} now results in
\begin{equation}
\eta = \frac{T_c}{T_{\mathrm{TRSB}}}\left( A \frac{ \langle \lvert f_a(\bo{p}) \rvert^2 \pazocal{I}[z(\bo{p})] \rangle_{\mathrm{FS}}}{ \langle \lvert f_a(\bo{p}) \rvert^2 \rangle_{\mathrm{FS}}} + B \frac{\langle \lvert f_b(\bo{p}) \rvert^2 \pazocal{I}[z(\bo{p})] \rangle_{\mathrm{FS}}}{ \langle \lvert f_a(\bo{p}) \rvert^2 \rangle_{\mathrm{FS}} } \right),
\label{eq:EtaNew}
\end{equation}
with $z(\bo{p}) = \frac{\Delta_0(T_{\mathrm{TRSB}})}{2k_B T_{\mathrm{TRSB}}} f_a(\bo{p})$.

\section{Microscopic Ginzburg--Landau theory}
\label{sec:GL}
In this appendix we summarize the derivation of the microscopic Ginzburg--Landau theory of Ref.~\onlinecite{WagnerEA21}. 

\subsection{Derivation}
\label{sec:Derivation}
\begin{figure}[t!bh]
	\centering
	\includegraphics[width=0.21\linewidth]{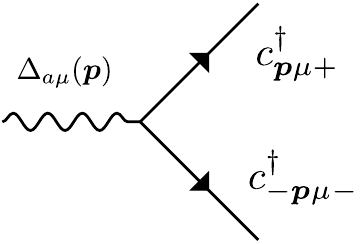}
\caption{Cooper pair operator in the (pseudo)spin singlet channel.}
	\label{fig:Cooper}
\end{figure}

\begin{figure}[t!bh]
	\centering
	\subfigure[]{\includegraphics[width=0.47\linewidth]{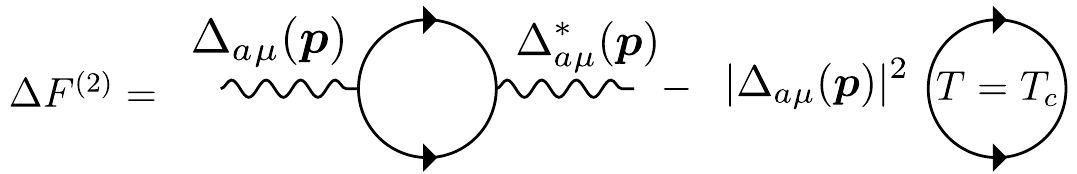}} \quad 
	\subfigure[]{\includegraphics[width=0.34\linewidth]{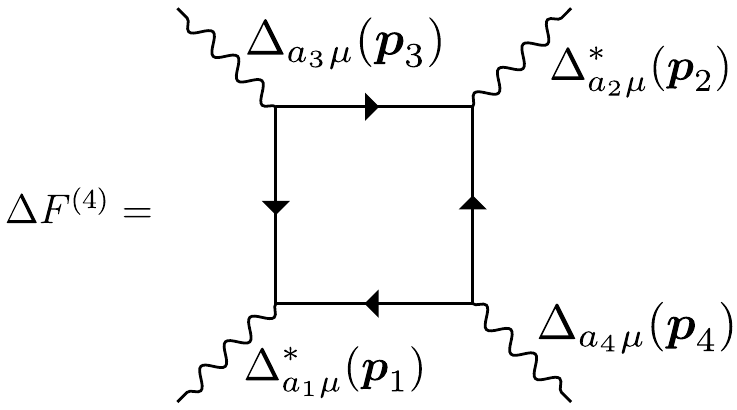}} 
\caption{The (a) second-order and (b) fourth-order diagrams contributing to the free energy of Eq.~\eqref{eq:LoopExpansion}. Corresponding algebraic expressions are given by Eqs.~\eqref{eq:SecondOrderF} and \eqref{eq:FourthOrderF}.}
	\label{fig:ContributionsF}
\end{figure}

We start with an electronic normal-state description,
\begin{equation}
    H_0 = \sum_{\mu, \sigma, \bo{p}} \xi_{\mu}(\bo{p}) c_{\bo{p}\mu\sigma}^{\dagger} c_{\bo{p} \mu \sigma},
    \label{eq:Hkin} 
\end{equation}
where, $\xi_{\mu}(\bo{p})$ is the dispersion of band $\mu$, and $\sigma = +, -$ denotes (pseudo)spin (with $\bar{\sigma}= -\sigma$). We here consider the pseudospin-singlet Cooper pairs and treat $H_{\text{SC}}$ as a perturbation to $H_0$ close to the critical temperature, where
\begin{equation}
\begin{aligned}
H_{\text{SC}} &= \sum_{\mu, \sigma, a, \bo{p} } \big[ \Delta_{a \mu}(\bo{p})  c_{\bo{p} \mu \sigma}^{\dagger} c_{-\bo{p} \mu  \bar{\sigma}}^{\dagger} + \text{h.c.} \big].
\end{aligned}
\label{eq:HSC}
\end{equation}
Here $\Delta_{a \mu}(\bo{p})$ is the order parameter on band $\mu$ with symmetry label (irreducible representation) $a$.  The sum over $\bo{p}$ runs over the Fermi surface sheet $\mu$ such that $\lvert \xi_{\mu} (\bo{p}) \rvert < \omega_c \sim k_BT$, where $\omega_c \ll W$ is the momentum cutoff and $W$ the bandwidth. Following the Ginzburg--Landau procedure, the free energy corrections to the normal state, $\Delta F = F_{\mathrm{SC}} - F_n$, are formally obtained by the Gibbs average~\cite{AbrikosovGorkovDzyal58, Gorkov59}
\begin{equation}
\Delta F = -T \ln\big\langle \pazocal{T}_{\tau} \exp\big( -\int_{0}^{\beta} \D\tau~H_{\text{SC}} (\tau)  \big) \big\rangle, \label{eq:LoopExpansion}
\end{equation}
where $\beta = 1/T$ (with $k_B = 1$), $\tau = -i t$ is imaginary time, and $\pazocal{T}_{\tau}$ is the time-ordering operator. In the loop expansion of $\Delta F$ we sum consistently connected diagrams, constructed from the Cooper operator of Fig.~\ref{fig:Cooper}, to a given order.
To quartic order we have the diagrams of Fig.~\ref{fig:ContributionsF}
The second- and quartic-order terms, $\Delta F =\Delta F^{(2)}+\Delta F^{(4)}$, are algebraically given by
\begin{align}
\label{eq:GLTheoryOne} \Delta F^{(2)} &=  {\sum_{\mu, a,\bo{p}}} \alpha_{a \mu}(\bo{p},T) \lvert \Delta_{a \mu}(\bo{p})\rvert^2  \\ 
\label{eq:GLTheoryTwo} \Delta F^{(4)} &= {\sum_{\mu, \lbrace a_i \rbrace, \lbrace \bo{p}_i \rbrace} } \beta_{\{a_i\} \mu}(\{\bo{p}_i\},T) \Delta^{*}_{a_1 \mu}(\bo{p}_1) \Delta^{*}_{a_2 \mu}(\bo{p}_2) \Delta_{a_3 \mu}(\bo{p}_3) \Delta_{a_4 \mu}(\bo{p}_4),
\\
\alpha_{a \mu}(\bo{p},T) &= - T \sum_{n} G_{\mu}(\bo{p}, \omega_n) G_{\mu}(-\bo{p}, -\omega_n)  + T_{c a}  \sum_{n} G_{\mu}(\bo{p}, \omega_n) G_{\mu}(-\bo{p}, -\omega_n) \rvert_{T = T_{c a}}, \label{eq:SecondOrderF} \\
\beta_{\{a_i\} \mu}(\{\bo{p}_i\}, T) &= \f{T}{2} g_{a_1 a_2 a_3 a_4} \delta_{\bo{p}_1 \bo{p}_3}\delta_{\bo{p}_1 \bo{p}_4} \delta_{\bo{p}_2 \bo{p}_3}\delta_{\bo{p}_2 \bo{p}_4} \sum_{n} G_{\mu}(\bo{p}_1, \omega_n)G_{\mu}(\bo{p}_2, \omega_n) G_{\mu}(-\bo{p}_3, -\omega_n)G_{\mu}(-\bo{p}_4, -\omega_n), \label{eq:FourthOrderF} \\
g_{a_1 a_2 a_3 a_4} &\coloneqq \delta_{a_1 a_3}\delta_{a_2 a_4}+\delta_{a_1 a_4}\delta_{a_2 a_3}+\delta_{a_1 a_2}\delta_{a_3 a_4}-2\delta_{a_1 a_2}\delta_{a_2 a_3}\delta_{a_3 a_4}. \label{eq:KroneckerCombination}
\end{align}
Above, the Green's functions $G_{\mu}(\bo{p}, \omega_n) = 1/(i\omega_n-\xi_{\mu}(\bo{p}))$ with fermionic Matsubara frequencies $\omega_n = \f{\pi}{\beta}(2n+1)$ for integer $n$ were introduced. Next, we factorize the order parameter as  $\Delta_{a \mu}(\bo{p})=\Delta_{0 a} f_{a \mu}(\bo{p})$, with $f_{a \mu}(\bo{p})$ being a (real) normalized form factor, and $\Delta_{0 a}$ is a (complex) variational parameter. The above equations simplify to
\begin{align}
\Delta F&=\Delta F^{(2)}+\Delta F^{(4)}= {\sum_{a}} \tilde\alpha_{a}(T,T_{c a}) \lvert \Delta_{0 a} \rvert^2  + {\sum_{\lbrace a_i \rbrace}} \tilde\beta_{\{a_i\}}(T_c) \Delta_{0 a_1}^{\ast} \Delta_{0 a_2}^{\ast} \Delta_{0 a_3} \Delta_{0 a_4},
\\
\tilde\alpha_{a}(T,T_{c a}) &= - V \sum_{\mu}\int \f{\D \bo{p}}{(2\pi)^d}~\Big( \f{\tanh\left[ \xi_{\mu}(\bo{p}) / (2T) \right] }{2\xi_{\mu}(\bo{p})} -\f{\tanh\left[ \xi_{\mu}(\bo{p}) / (2T_{ca}) \right]}{2\xi_{\mu}(\bo{p})} \Big) f^2_{a \mu}(\bo{p}), \label{eq:AppSecondOrderF} \\
\tilde\beta_{\{a_i\}}(T) &= g_{a_1 a_2 a_3 a_4} \f{V}{2T^3} \sum_\mu \int \f{\D \bo{p}}{(2\pi)^d}~h( \xi_{\mu}(\bo{p})/ T )f_{a_1 \mu}(\bo{p}) f_{a_2 \mu}(\bo{p}) f_{a_3 \mu}(\bo{p}) f_{a_4 \mu}(\bo{p}), \label{eq:AppFourthOrderF}
\end{align}
after performing the Matsubara sums. Above, $V$ is the unit cell volume, the momentum dependent order parameter components are without loss of generality assumed real, and
\begin{equation}
h(x) \coloneqq \frac{\sinh{x}-x}{4x^3(1+\cosh{x})}.
\label{eq:hfunc}
\end{equation}
\subsection{Minimization in the two-component case}
\label{sec:TwoComponent}
In the two-component case of Eq.~\eqref{eq:OPansatz} the free energy of Eq.~\eqref{eq:FreeEnergy} is minimized over $X(T)$ and $\Delta_0(T)$. When redefining $\tilde{\alpha}_j \to \alpha_j$ and $\tilde{\beta}_j \to \beta_j$ for notational convenience, we get
\begin{equation}
    X^2(T) =  \begin{cases}
        \frac{\alpha_a(T) \beta_{ab} - \alpha_b(T) \beta_{a}}{\alpha_b(T) \beta_{ab} -\alpha_a(T)\beta_{b}} & \text{for } T<T_\textrm{TRSB}\\
        0 & \text{for } T>T_\textrm{TRSB}
        \end{cases},
\end{equation}
and
\begin{equation}
    \Delta_0^2(T) = \begin{cases}
        -\frac{1}{2}\frac{\alpha_b(T) \beta_{ab}-\beta_{b} \alpha_a(T)}{\beta_{ab}^2-\beta_{b} \beta_{a}} & \text{for } T<T_\textrm{TRSB}\\
        -\frac{\alpha_a(T)}{2\beta_{a}} & \text{for } T_\textrm{TRSB}<T<T_c\\
        0 & \text{for } T>T_c
        \end{cases}.
\end{equation}
After making the approximations ${\alpha}_a = {\alpha}_a^0\left( \frac{T}{T_{c1}} - 1 \right)$ and ${\alpha}_b = {\alpha}_b^0\left( \frac{T}{T_{c2}} - 1 \right)$, with $T_{c2} < T_{c1}$, the temperature $T_{\mathrm{TRSB}}$ is given by
\begin{equation}
    T_\textrm{TRSB}(T_{c1}, T_{c2}) = T_{c2}\frac{1-\frac{\beta_{ab}\alpha_a^0}{\beta_a\alpha_b^0}}{1-\frac{T_{c2}}{T_{c1}}\frac{\beta_{ab}\alpha_a^0}{\beta_a\alpha_b^0}}<T_{c2}.
    \label{eq:TTRSB}
\end{equation}
For convenience we also introduce the temperature
\begin{equation}
    T_{\ast}(T_{c1}, T_{c2}) = T_{c1}\frac{1-\frac{\beta_{ab}\alpha_b^0}{\beta_b\alpha_a^0}}{1-\frac{T_{c1}}{T_{c2}}\frac{\beta_{ab}\alpha_b^0}{\beta_b\alpha_a^0}}>T_{c1},
    \label{eq:Tstar}
\end{equation}
such that
\begin{equation}
     \Delta_0^2(T<T_\textrm{TRSB})=- \frac{1}{2}\frac{\beta_b\alpha_a^0-\beta_{ab}\alpha_b^0}{\beta_{ab}^2-\beta_a\beta_b}\left(1-\frac{T}{T_{\ast}}\right)=\Delta_0^2(0)\left(1-\frac{T}{T_{\ast}}\right)
\end{equation}
and
\begin{equation}
    X^2(T<T_\textrm{TRSB})=\frac{\beta_{ab}\alpha_a^0-\beta_a\alpha_b^0}{\beta_{ab}\alpha_b^0-\alpha_a^0\beta_b}\frac{1-\frac{T}{T_\textrm{TRSB}}}{1-\frac{T}{T_*}}=X^2(0)\frac{1-\frac{T}{T_\textrm{TRSB}}}{1-\frac{T}{T_*}}.
\end{equation}
\section{Three-band tight binding model for $\mathrm{Sr}_2\mathrm{RuO}_4$}
\label{sec:Bandstructure}
We employ an effective 2D three-band model for the band structure of Sr$_2$RuO$_4$,
\begin{equation}
H_0 = \sum_{\bo{k}, s} \bo{\psi}^{\dagger}_s(\bo{k}) h_s(\bo{k}) \bo{\psi}_s(\bo{k}),
\label{eq:TBHamiltonian}
\end{equation}
where $\bo{\psi}_s(\bo{k}) = [c_{ xz, s}(\bo{k}), \hspace{1mm} c_{yz, s}(\bo{k}), \hspace{1mm} c_{xy, -s}(\bo{k})]^T$ and where  $s \in \lbrace \uparrow, \downarrow \rbrace$ denotes spin and $a\in\{xz,yz,xy\}$ denotes the $d$-orbitals of the Ru atoms, which are relevant close to the Fermi energy. For $h_s(\bo{k})$ we take
\begin{equation}
h_s(\bo{k}) = \begin{pmatrix}
\varepsilon_{xz}(\bo{k}) & g(\bo{k}) - i s\lambda &  i\lambda \\
g(\bo{k})+ i s\lambda & \varepsilon_{yz}(\bo{k}) &  - s\lambda \\ - i\lambda & - s\lambda & \varepsilon_{xy}(\bo{k})
\end{pmatrix},
\label{eq:H0}
\end{equation}
where spin-orbit coupling is parametrized by $\lambda$, and the above energies are given by
\begin{align}
\varepsilon_{xz}(\bo{k}) &= - 2t_1(1-p) \cos{k_x} - 2t_2(1+\nu p) \cos(k_y)  - \mu, \label{eq:xzhopping} \\
\varepsilon_{yz}(\bo{k}) &= - 2t_1(1+\nu p) \cos{k_y} - 2t_2(1- p) \cos(k_x)  - \mu, \label{eq:yzhopping} \\
\varepsilon_{xy}(\bo{k}) &= - 2t_3 \left[ (1-p)\cos{k_x} + (1+\nu p)\cos{k_y} \right] - 4t_4 \cos{k_x} \cos{k_y} \nonumber \\ 
&\hspace{20pt} - 2t_5\left[ (1-p)\cos(2k_x) + (1+\nu p) \cos(2k_y) \right] - \mu , \label{eq:xyhopping} \\
g(\bo{k}) &= -4 t' \sin{k_x} \sin(k_y).
\end{align}
Here (100) strain is parameterized by the deformation parameter $p$ and the Poisson's ratio $\nu$. The tight-binding parameters are summarized in Table~\ref{tab:HoppingParameters} and are consistent with ARPES measurements~\cite{ZabolotnyyEA13}.
\begin{table}
\centering
\caption{Tight-binding parameters for Eqs.~\eqref{eq:H0} consistent with ARPES measurements~\cite{ZabolotnyyEA13}, and Poisson's ratio $\nu$ applicable to (100) strain taken from Ref.~\cite{BarberEA19}.}
\begin{tabular}{p{2.5cm} p{1.3cm} p{1.3cm} p{1.3cm} p{1.3cm} p{1.3cm} p{1.3cm} p{1.3cm} p{1.3cm} p{1.3cm}}  
\toprule
Parameter & $t_1$ & $t_2$ & $t_3$ & $t_4$ & $t_5$ & $t'$ & $\mu$ & $\lambda$ & $\nu$ \\ \hline
Value [meV] & $88$ & $9$ & $80$ & $40$ & $5$ & $4$ & $109$ & $45$ & $0.508$  \\ \bottomrule
\label{tab:HoppingParameters}
\end{tabular}
\end{table}
\section{Additional results}
\label{sec:MorePlots}
\begin{figure}[t!bh]
	\centering
	\includegraphics[width=0.65\linewidth]{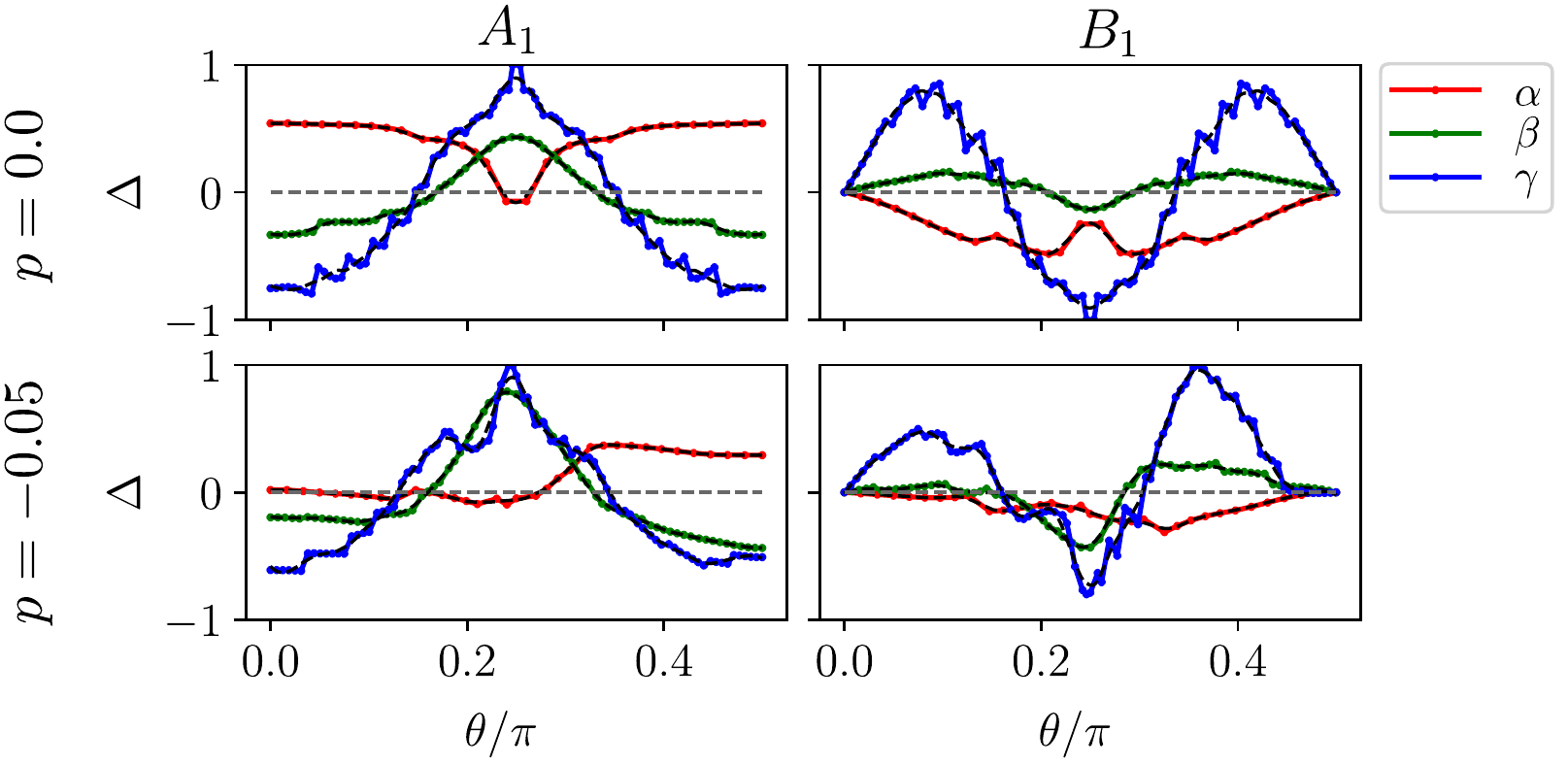}
\caption{Order parameters from Fig.~\ref{fig:OrderParameters} labelled by the respective irreps of $D_2$, as a function of polar angle (on the $\alpha$ band the polar angle is measured from $(\pi,\pi)$). The coloured data points are the raw data from Ref.~\onlinecite{RomerEA21}, and the black dashed lines show the data passed through a Savitzky--Golay filter to reduce finite size effects. The filtered order parameters were used in the calculations of Sec.~\ref{sec:Results}.}
	\label{fig:OrderParameterspolar}
\end{figure}

\begin{figure}[t!bh]
	\centering
	\includegraphics[width=0.65\linewidth]{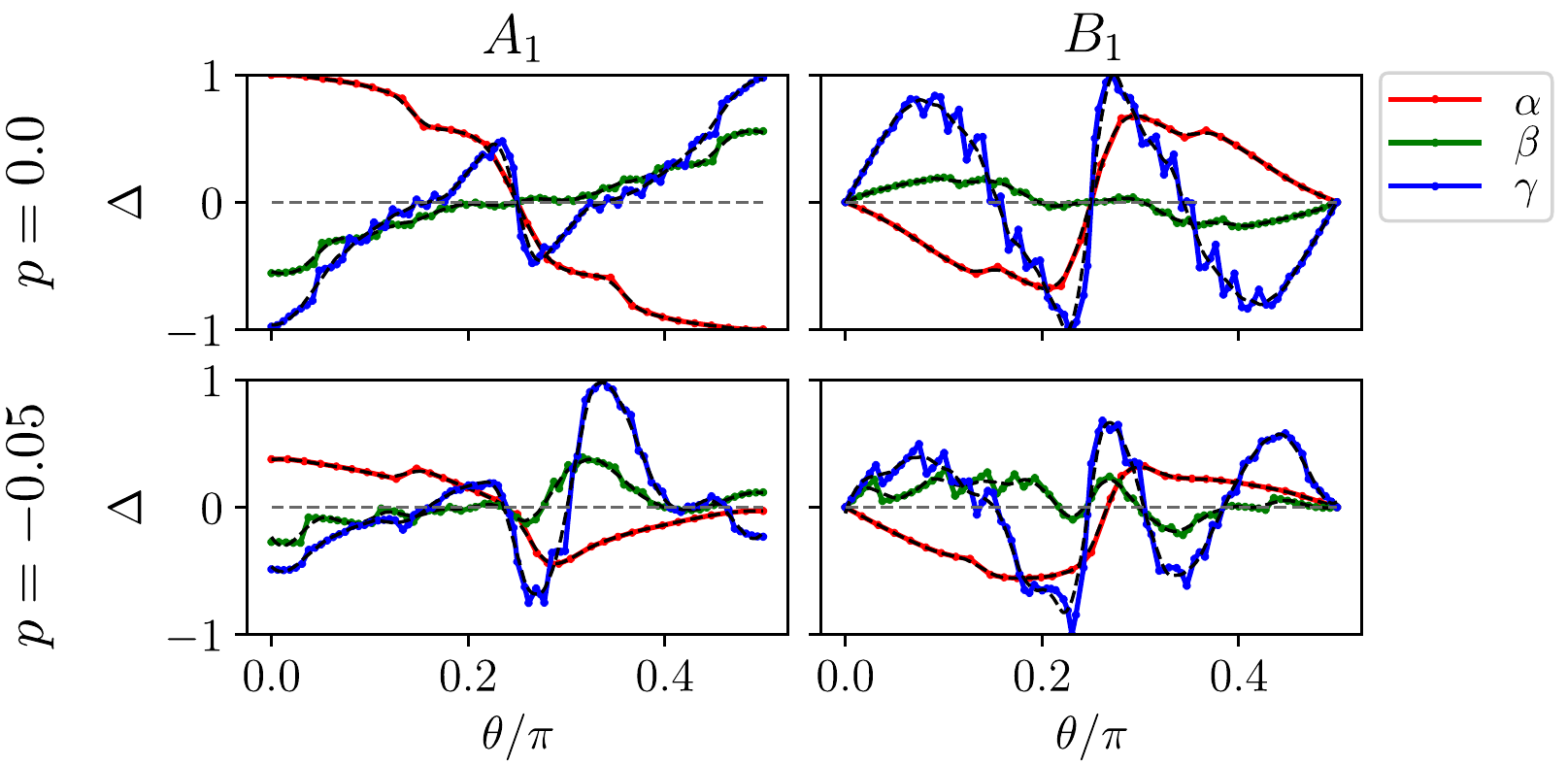}
\caption{Order parameters from Fig.~\ref{fig:OrderParameters2} labelled by the respective irreps of $D_2$. See the caption of Fig.~\ref{fig:OrderParameterspolar} for further details.}
	\label{fig:OrderParameterspolar2}
\end{figure}

\begin{figure}[t!bh]
	\centering
	\includegraphics[width=0.8\linewidth]{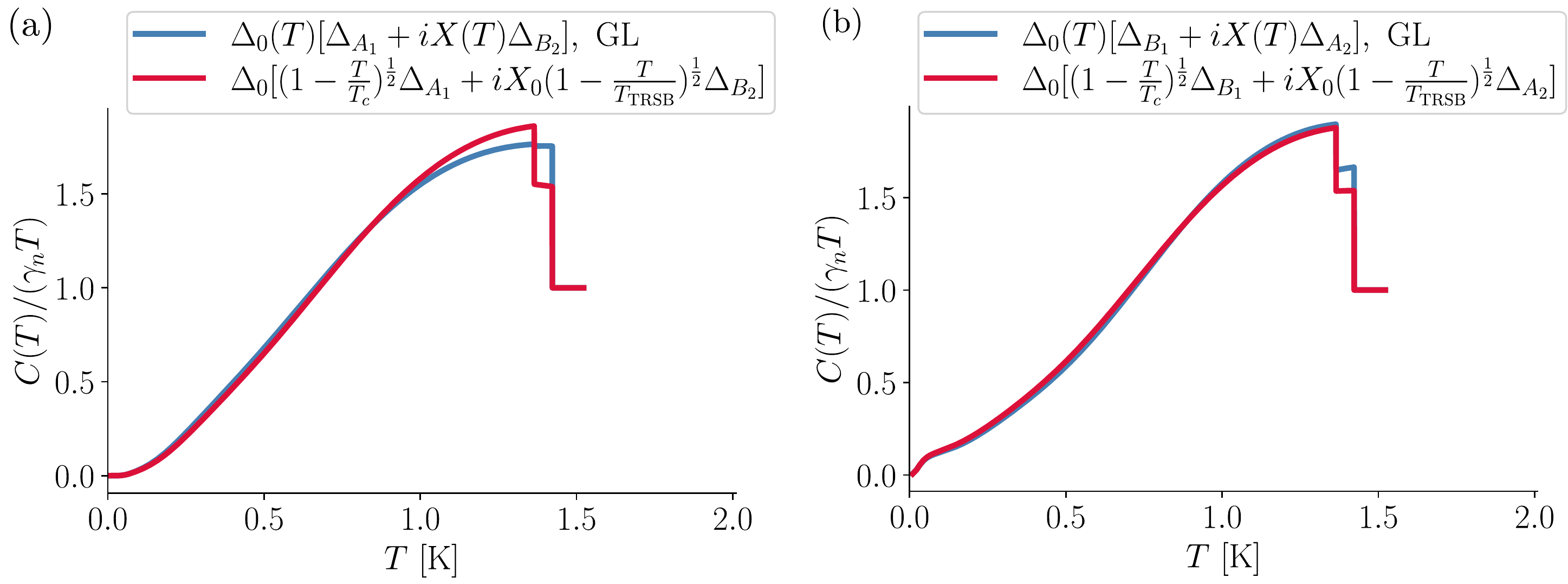} 
\caption{Comparison of the heat capacity per temperature and normal state value for the (a) $s'+id_{xy}$ and (b) $d_{x^2-y^2}+ig_{xy(x^2-y^2)}$ order parameters from Ref.~\onlinecite{RomerEA21} at zero strain using two different temperature dependencies. The blue lines show results using the Ginzburg--Landau solution, while the red lines use simple square root temperature dependencies and a constant $X_0 = 0.8$. In both cases it is seen that the non-analyticity of $\Delta_0(T=T_{\mathrm{TRSB}})$ reduces the second jump.}
	\label{fig:HeatCapacityComparison}
\end{figure}

In Figs.~\ref{fig:OrderParameterspolar} and \ref{fig:OrderParameterspolar2} we show supplementary plots of the order parameter components from Figs.~\ref{fig:OrderParameters} and \ref{fig:OrderParameters2} as a function of polar angle.
In Fig.~\ref{fig:HeatCapacityComparison} we compare the heat capacity resulting from the Ginzburg--Landau theory, i.e.,~with an order parameter of the form
\begin{equation}
    \Delta(T, \bo{p}) = \Delta_0(T) \left[ f_{a}(\bo{p}) + i X(T) f_{b}(\bo{p}) \right],
\end{equation}
with $X(T)$ and $\Delta_0(T)$ minimizing Eq.~\eqref{eq:FreeEnergy}, to the results of using an order parameter of the form (cf.~Ref.~\onlinecite{KivelsonEA20})
\begin{equation}
    \Delta(T, \bo{p}) = \Delta_0 \left[ \sqrt{1-\frac{T}{T_c}} f_{a}(\bo{p}) + i X_0 \sqrt{1-\frac{T}{T_{\mathrm{TRSB}}}} f_{b}(\bo{p}) \right],
\end{equation}
in both cases with the momentum dependent components from Fig.~\ref{fig:OrderParameterspolar} and \ref{fig:OrderParameterspolar2} and in the latter case with fixed $X_0 = 0.8$. In the case of both $s'+id_{xy}$ (panel (a)) and $d_{x^2-y^2}+ig_{xy(x^2-y^2)}$ (panel (b)) it is seen that the effect of the non-analyticity of $\Delta_0(T)$ at $T = T_{\mathrm{TRSB}}$ is a reduction of the second jump. In the case of panel (a) the second jump is almost vanishing for the Ginzburg--Landau solution.

\begin{figure}[t!bh]
	\centering
	\includegraphics[width=0.8\linewidth]{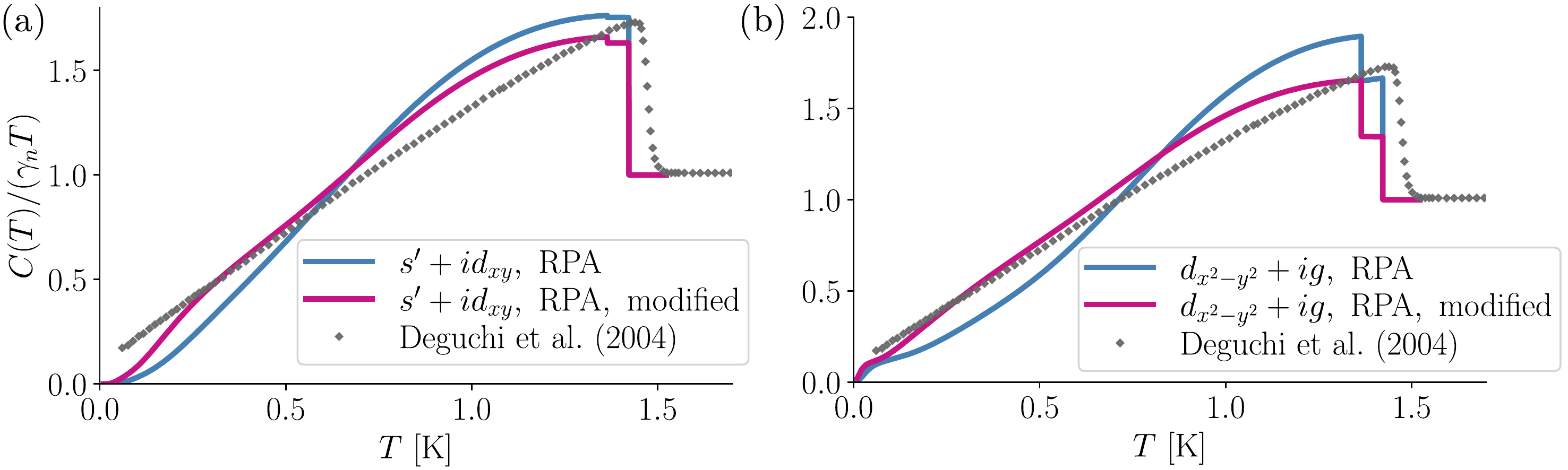} 
\caption{Comparison of the heat capacity per temperature and normal state value at zero strain using the Ginzburg--Landau solution with form factors obtained from microscopic RPA calculations~\cite{RomerEA21} (blue lines) to a case where the band-resolved gap magnitudes have been redistributed to obtain a better fit (purple lines) with the experimental data of Ref.~\onlinecite{DeguchiEA04} (gray diamonds). Panel (a): The $s'+id_{xy}$ scenario. Here we redistributed $\Delta_{\alpha}/\Delta_{\gamma} = 0.7$ and $\Delta_{\beta}/\Delta_{\gamma} = 0.6$ as compared to the bare RPA form factors to obtain the purple line. Panel (b): The $d_{x^2-y^2}+ig_{xy(x^2-y^2)}$ scenario, where $\Delta_{\alpha}/\Delta_{\gamma} = 0.4$ and $\Delta_{\beta}/\Delta_{\gamma} = 0.5$ characterizes the purple line.}
	\label{fig:HeatCapacity_RPAmodified}
\end{figure}

In Fig.~\ref{fig:HeatCapacity_RPAmodified} we demonstrate that the match with the experimental heat capacity data of Ref.~\onlinecite{DeguchiEA04} can be somewhat improved by redistributing the band-resolved gap magnitudes compared to those obtained with the microscopic RPA calculations of Ref.~\onlinecite{RomerEA21}. A past study showed that by tuning interaction parameters of the theory it is possible to obtain a match with experimental data, albeit the study focused on an $E_u$ (of $D_{4h}$) phase~\cite{NomuraEA02}.

\begin{figure}[t!bh]
	\centering
	\includegraphics[width=0.6\linewidth]{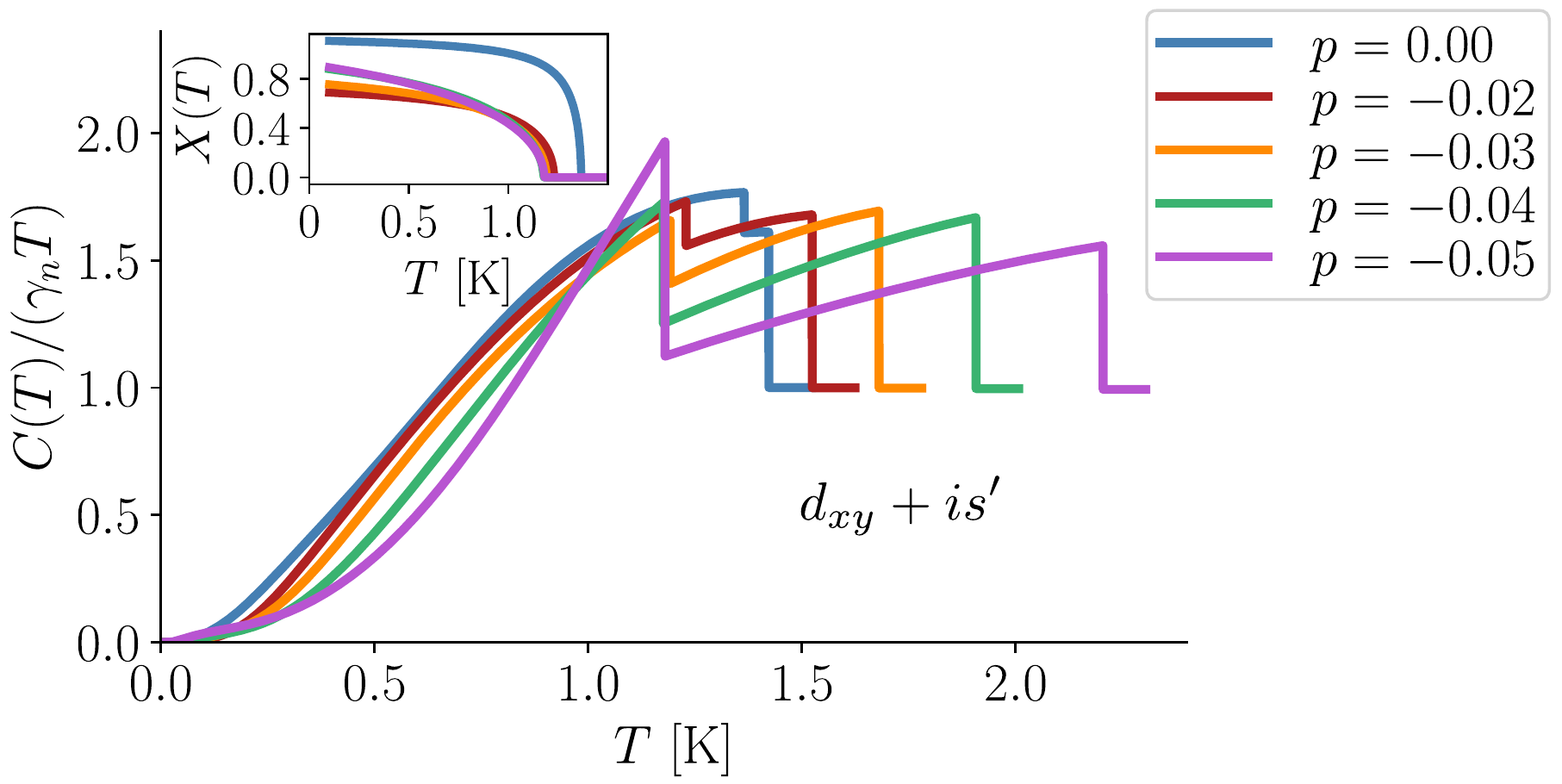}
\caption{Supplement to Fig.~\ref{fig:HeatCapacity}(a): Heat capacity for the $d_{xy}+is'$ order parameter. We used the order parameter ansatz of Eq.~\eqref{eq:OPansatz} with $\Delta_0(T)$ and $X(T)$ (inset) resulting from minimization of the free energy of Eq.~\eqref{eq:FreeEnergy}, and with the momentum dependent form factors of Fig.~\ref{fig:OrderParameters}.}
	\label{fig:HeatCapacityFlipped}
\end{figure}

In Fig.~\ref{fig:HeatCapacityFlipped} we provide supplementary results of the heat capacity for the same band structure, transition temperatures, and order parameter components as in Sec.~\ref{sec:Results}, but for the reversed combination of orders, i.e.,~$d_{xy}+is'$ (strictly $B_1 + iA_1$ of $D_2$). The resulting second heat capacity jump is found to be $\eta = 0.260,~ 0.255,~0.356,~ 0.707,~1.496$ for $p = 0.0, -0.02, -0.03, -0.04, -0.05$, respectively.
Not only does this rule out this specific $d_{xy}+is'$ order parameter as a viable candidate order for SRO~\cite{LiEA19}, but it verifies the principles of Sec.~\ref{sec:Parameters}. Physically, the $B_1$ component is nodal along $k_x,~k_y = 0$, so if this component is the primary one, it leaves quasiparticles to be gapped out by the $A_1$ component (which is large at these points), resulting in a sizable second jump. This is qualitatively different from the case where the $A_1$ is the primary component. It should also be noted that the nodal regions of the $B_1$ component on the $\gamma$ sheet grow with strain; see Fig.~\ref{fig:OrderParameters}.

\begin{figure}[t!bh]
	\centering
	\includegraphics[width=0.8\linewidth]{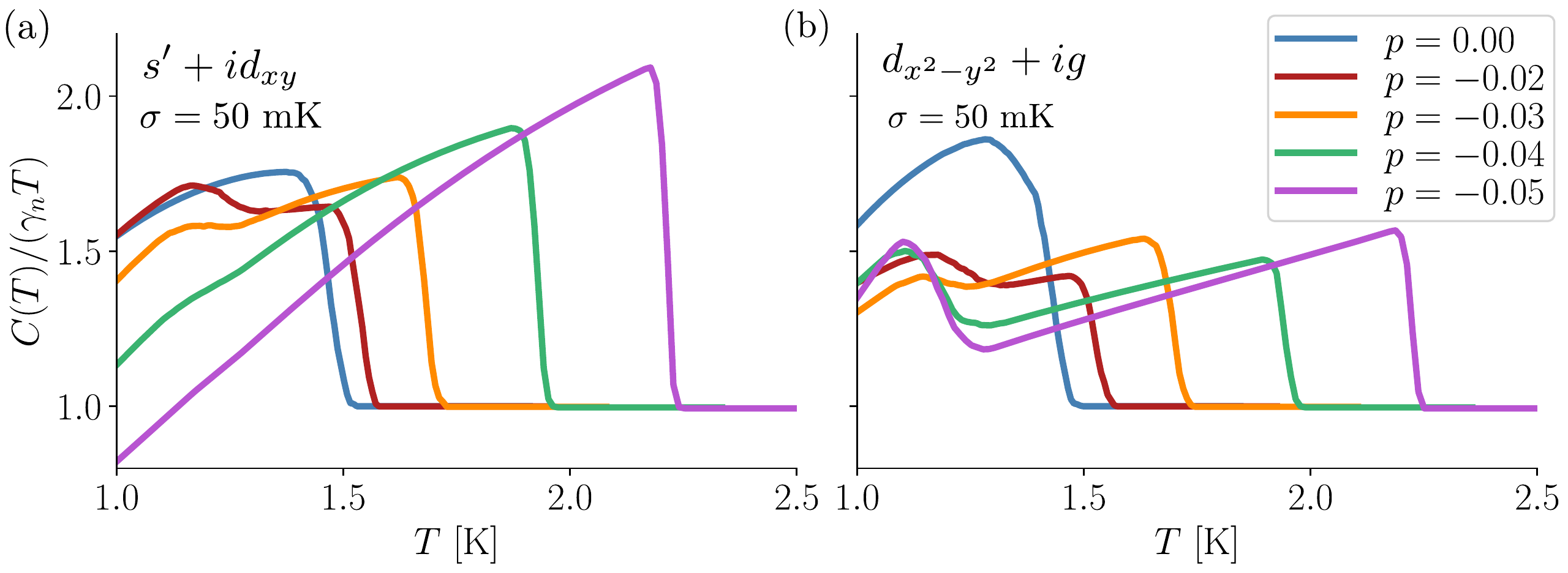}
\caption{Simulated effect of smearing with a transition width of $\sigma = 50~\mathrm{mK}$ in the (a) $s'+id_{xy}$ and (b) $d_{x^2-y^2}+ig_{xy(x^2-y^2)}$ order parameter scenario. See the caption of Fig.~\ref{fig:HeatCapacitySmear} for further details.}
	\label{fig:HeatCapacitySmearSupp}
\end{figure}

Finally, in Fig.~\ref{fig:HeatCapacitySmearSupp} we supplement Fig.~\ref{fig:HeatCapacitySmear} with the simulated effect of a modest smearing of $\sigma = 50~\mathrm{mK}$, which is in the lower end of the scale of the experimentally relevant transition widths~\cite{LiEA19}. In this case, the second jump is resolvable for both the $s'+id_{xy}$ and the $d_{x^2-y^2}+ig_{xy(x^2-y^2)}$ order parameters, at least for some strain values. 

\end{appendix}
\end{document}